\documentclass[12pt,letterpaper,usenames,dvipsnames]{article}
\usepackage{Auxiliary/jheppub}
\allowdisplaybreaks[4]
\usepackage{tcolorbox}
\usepackage{comment}
\usepackage{dsfont}
\usepackage{amsmath,amssymb,amsfonts,amsthm,amscd,setspace}
\usepackage{bm} 
\usepackage{bbm} 
\usepackage{mathbbol} 

\usepackage{graphicx,tikz}
\usepackage[framemethod=TikZ]{mdframed} 
\usepackage{tikz-cd} 
\usetikzlibrary{arrows,snakes,shapes.arrows,decorations.markings}
     \tikzset{>=triangle 90}
     \tikzstyle{gr}=[draw,circle,green!50!black,fill=green!50!black,scale=.6]
     \tikzstyle{Bl}=[draw,circle,blue,scale=.6]
     \tikzstyle{R}=[draw,circle,fill=red,scale=.6]
     \tikzstyle{bl}=[draw,circle,fill=black,scale=.35]
     \tikzstyle{bbc}=[draw,circle,fill=black,scale=.75]
     \tikzstyle{bbcs}=[draw,circle,fill=black,scale=.5]
     \tikzstyle{rc}=[circle,fill=red,scale=.6]
     \tikzstyle{wc}=[draw,circle,scale=.75]

\usepackage{multicol}
\usepackage[export]{adjustbox}
\usepackage{rotating} 
\usepackage{colortbl}
\usepackage{floatrow}

\usepackage{eufrak} 
\usepackage{mathrsfs} 
\usepackage{enumitem}
\usepackage{tcolorbox,array,hhline,arydshln,multirow,floatrow,subfig}

\usepackage{xcolor}
\def\red#1{{\color{red}{#1}}}
\def\blue#1{{\color{blue}{#1}}}
\def\green#1{{\color{black!30!green}{#1}}}

\def\yellow#1{{\color{black!25!yellow!70!red}{#1}}}

\def\nn{\nonumber}

\def\bM{\begin{matrix}}
\def\eM{\end{matrix}}
\newcommand{\bpm}{\begin{pmatrix}}
\newcommand{\epm}{\end{pmatrix}}
\newcommand{\bsm}{\begin{smallmatrix}}
\newcommand{\esm}{\end{smallmatrix}}
\newcommand{\bspm}{\left(\begin{smallmatrix}}
\newcommand{\espm}{\end{smallmatrix}\right)}

\def\tilde{\widetilde}

\def\^{\wedge}
\def\del{{\partial}}

\def\Im{{\rm Im}}

\def\GL{{\rm GL}}

\def\SO{{\rm SO}}

\def\SU{{\rm SU}}
\def\su{\mathfrak{su}}
\def\Sp{{\rm Sp}}

\def\U{{\rm U}}

\def\C{\mathbb{C}}

\def\ff{{\mathfrak f}}

\def\I{\mathbb{I}}

\def\cN{{\mathcal N}}
\def\cO{{\mathcal O}}

\def\P{\mathbb{P}}

\def\cV{{\mathcal V}}

\def\Z{\mathbb{Z}}
\def\a{{\alpha}}

\def\D{{\Delta}}

\def\z{{\zeta}}

\def\l{{\lambda}}
\def\L{{\Lambda}}

\def\t{{\tau}}

\def\w{{\omega}}

\def\del{{\partial}}

\def\^{\wedge}
\def\I{\mathds{1}}

\def\Im{{\rm Im}}

\def\U{{\rm U}}
\def\SU{{\rm SU}}
\def\SO{{\rm SO}}

\def\GL{{\rm GL}} 
\def\Sp{{\rm Sp}}

\def\ef{\mathfrak{e}}
\def\ff{\mathfrak{f}}
\def\gf{\mathfrak{g}}

\def\rSf{\red{\mathfrak{S}}}

\def\sof{\mathfrak{so}}
\def\spf{\mathfrak{sp}}
\def\suf{\mathfrak{su}}

\def\uf{\mathfrak{u}}

\def\cN{{\mathcal N}}
\def\cO{{\mathcal O}}

\def\cV{{\mathcal V}}

\def\C{\mathbb{C}}

\def\P{\mathbb{P}}

\def\Z{\mathbb{Z}}

\def\beq{\begin{equation}}
\def\eeq{\end{equation}}
\def\nn{\nonumber}
\newcommand{\bpmat}{\begin{pmatrix}}
\newcommand{\epmat}{\end{pmatrix}}
\newcommand{\bsmat}{\begin{smallmatrix}}
\newcommand{\esmat}{\end{smallmatrix}}
\newfloatcommand{capbtabbox}{table}[][\FBwidth]

\def\ccg{\cellcolor{green!07}}
\def\ccgy{\cellcolor{black!25!green!50!yellow!15}}

\def\rcb{\rowcolor{blue!07}}

\def\rcy{\rowcolor{black!25!yellow!10}}

\title{The rank 2 classification problem I:\\ scale invariant geometries}
\author[1]{Philip C. Argyres}
\author[2]{Mario Martone}
\affiliation[1]{University of Cincinnati, Physics Department, PO Box 210011, Cincinnati OH 45221}
\affiliation[2]{Dept.\ of Mathematics, King’s College London, The Strand, London WC2R 2LS, UK}
\emailAdd{philip.argyres@gmail.com}
\emailAdd{mario.martone@kcl.ac.uk}

\abstract{In this first of a series of three papers we outline an approach to classifying 4d $\cN{=}2$ superconformal field theories at rank 2.  
The classification of allowed scale invariant $\cN=2$ Coulomb branch geometries of dimension (or rank) greater than one is a famous open problem whose solution will greatly constrain the space of $\cN{=}2$ superconformal field theories. 
At rank 2 the problem is equivalent to finding all possible genus 2 Seiberg-Witten curves and 1-forms satisfying a special K\"ahler condition.
This is tractable because regular genus 2 Riemann surfaces can be uniformly described as binary-sextic plane curves, and the Seiberg-Witten curves are families of such curves varying meromorphically over the two-dimensional base. 
There are also solutions consisting of families of degenerate genus-2 Riemann surfaces given by a bouquet of two elliptic curves which are described by a different set of curves. 
In this paper we set up and carry out the analysis of the generic case, \emph{i.e.}, those whose typical fiber is a regular genus-2 Riemann surface with no extended automorphism, and find the complete answer for polynomial coefficients.} 
  
\begin{document}
\maketitle

\section{Introduction}

Our ultimate goal is to classify all possible rank 2 scale-invariant special K\"ahler (SK) Coulomb branch (CB) geometries.%
\footnote{The rank is the complex dimension of the CB.  Also, the term ``Coulomb branch geometry'' denotes a set of further constraints beyond SK-ness which come from physical considerations, and which will be spelled out later.}
This goal is interesting since this list of geometries constrains the set of possible rank 2 4d $\cN{=}2$ SCFTs.
At the moment, it is not even known whether there is a finite set of such theories,%
\footnote{A continuum of SCFTs connected by exactly marginal deformations will count as a single SCFT in this enumeration.}
though there is every indication and expectation that the number is finite and even relatively small, say, less than 100 \cite{Martone:2021ixp} (so far, including those discovered with the analysis presented here and discussed in \cite{Argyres:2022puv}, only 73 are known to exist).

In the rank 1 case, there are only 7 non-trivial scale-invariant CB geometries, famously given by a subset of the Kodaira classification of degenerations of an elliptic fibre over $\C$ \cite{Minahan:1996fg}.%
\footnote{They give rise to 27 distinct geometries upon deformation \cite{Argyres:2015ffa, Argyres:2015gha, Argyres:2016xua, Argyres:2016xmc}, all but 2 of which have known corresponding SCFTs.}
The elliptic fibre arises as the fibre of the genus 1 Seiberg-Witten (SW) curve \cite{Seiberg:1994rs, Seiberg:1994aj} describing the rank 1 SK CB geometry. 
This result only holds if the CB has no complex singularities \cite{Bourget:2018ond, Argyres:2018wxu}, and more options are possible otherwise \cite{Argyres:2017tmj}. 
We will not consider here the most general case and henceforth assume that complex singularities are absent. 

By contrast, a general rank $r$ CB geometry can be written in terms of complex geometry as a family of rank $r$ polarized abelian varieties varying holomorphically over the CB together with a holomorphic symplectic form on the total space of the family with vanishing restriction to the fibres of the family \cite{Donagi:1995cf}.  
General abelian varieties, though algebraic, are difficult to describe in a uniform way algebraically.  
However, at rank 2 all principally polarized abelian varieties arise as the Jacobian variety of a genus 2 Riemann surface (RS) or of a singular limit of such a surface.
Furthermore, all non-singular genus 2 RSs can be written as a suitable projectivization of binary sextic plane curves. 
The SW curve is then a family of such genus-2 curves varying meromorphically over $\C^2$.

So, at least for the principally polarized case and CBs with no complex singularities, there is a uniform and relatively simple algebraic setting in which all rank 2 CB geometries can be explored.
Concretely, take $(u,v) \in \C^2$ to be complex coordinates on the rank 2 CB.
Then the SW curve is
\begin{align}\label{SWcurve}
    y^2 &= c(x;u,v), 
\end{align}
where $c$ is a degree 6 polynomial in $x$ which may depend locally meromorphically on $(u,v)$.
The coefficients of \eqref{SWcurve} must also satisfy extra conditions following from $\cN{=}2$ supersymmetry, scale-invariance, and unitarity, reviewed below.
These conditions are algebraically complicated and very constraining.
Upon making one further simplifying technical assumption, described below, we will show that these conditions admit only fourteen solutions which can be interpreted as CBs of $\cN=2$ SCFTs.
These solutions are listed in table \ref{tab:knownSW}.
The $\mathfrak{S}_a$ stratification columns are explained in section \ref{strat}.

\begin{table}[t!]
\begin{adjustbox}{center,max width=1\textwidth}
$\def\arraystretch{1.0}
\begin{array}{c|c:c:c|c}
\multicolumn{5}{c}{\Large \textsc{CBs given by polynomial families of genus 2 RSs}}\\
\hline
\hline
 \D_{u,v}&\ \red{\mathfrak{S}}_{\rm knot} \ \, &\quad \red{\mathfrak{S}}_u \quad\, &\quad \red{\mathfrak{S}}_v \quad\,
&\textrm{Form of the curve in canonical frame}
\\
\cdashline{1-5}

 \{6,8\}
&I_1&\varnothing & I_6^*
&\ccg y^2=x\, (u\, x-v)\,(x^4+u\, x-v)\\[1mm]

\{4,10\}
&I_1&\varnothing & II^*
&y^2=x^5+ (u\, x-v)^3\\[1mm]

\{4,6\}
&I_1&I_2 & I_4^*
&\ccg y^2=x\, (u\,x-v)\, (x^3 +u\, x-v)\\[1mm]
 \{4,5\}

&I_1&\varnothing & I_{10}
&\ccg y^2=(u\, x-v)(x^5+u\, x-v) \\[1mm]

 \{3,5\}

&I_1&\varnothing & I_3^*
&y^2=x(x^5+(u\, x-v)^2) \\[1mm]
 \{3,4\}

&I_1& I_2 & I_8
&\ccg y^2=(u\, x-v)(x^4+u\, x-v)\\[1mm]

\rcy \{2,4\}

&(I_1)^2&\varnothing & I_2^*
& y^2=x(x^4+\t\,x^2\,(u\, x-v)+(u\,x-v)^2)\\[1mm]

\rcy \{2,3\} 

&(I_1)^2&\varnothing & I_6
&y^2=x^6+\t x^3\, (u\,x-v)+ (u\, x-v)^2\\[1mm]

\rcy \{2,2\}

&(I_2)^3&I_2 & I_2
& \ccgy y^2=(u\, x-v)(x^5 + \t_1 x^3 + \t_2 x^2 + \t_3 x + \t_4)\\[1mm]

\{\frac32,\frac52\} 

&I_1&\varnothing & I_5
&y^2=x^5+(u\,x-v)^2\\[1mm]

\{\frac43,\frac53\}

&I_1&\varnothing & I_2 
&y^2=x(x^5+u\,x-v)\\[1mm]

\{\frac65,\frac85\}

&I_1&\varnothing & I_2
&y^2=x(x^4+u\,x-v)\\[1mm]

\{\frac54,\frac32\} 
&I_1&\varnothing & \varnothing
&y^2=x^6+u\,x-v\\[1mm]

\{\frac87,\frac{10}7\} 

&I_1&\varnothing & \varnothing
&y^2=x^5+u\,x-v\\[.5mm]

\hline
\hline

\end{array}$
\caption{\label{tab:knownSW}
All rank 2 scale invariant CB geometries described by genus 2 SW curves with polynomial coefficients. 
The first column gives the CB scaling dimensions.
The next three columns list the monodromies around each co-dimension one stratum. 
$\varnothing$ means that the corresponding stratum is not actually part of the singular locus and the power means that there are multiple components of that given form. 
The last column gives the binary-sextic plane curve; in green are the solutions which are new to this paper, and in yellow the solutions which correspond to Lagrangian theories.
}
\end{adjustbox}
\end{table}

The $\{\D_u,\D_v\}=\{2,2\}$ solution shown in the table is new, and depends on four arbitrary couplings, instead of the two expected from the gauge SCFT.
Its CB stratification coincides with that expected from the $\SU(2)\times\SU(2)$ superconformal gauge theory with a bi-fundamental hypermultiplet as well as two fundamental hypermultiplets for each gauge factor, which has an $\SO(4)^2\times\Sp(2) \simeq \SU(2)^5$ flavor symmetry.
The known curve for this theory \cite{Argyres:1999fc} is a specialization of the one shown in the table, found, \emph{e.g.}, by setting $\t_3=1$ and $\t_4=0$.
The meaning of the extra parameters in this solution is unclear;  see the discussion of this case in appendix \ref{appA22}

\begin{center}
\rule[1mm]{2cm}{.4pt}\hspace{1cm}$\circ$\hspace{1cm} \rule[1mm]{2cm}{.4pt}
\end{center}

The remainder of this introduction is devoted to outlining how a systematic search for scale invariant rank 2 CB geometries can be carried out using binary-sextic SW plane curves.
This setting was used in a previous attempt \cite{Argyres:2005pp, Argyres:2005wx} (in which one of the authors was involved) at constructing rank 2 geometries.
In particular, those papers fixed the reparameterization freedom of the curve \eqref{SWcurve} by choosing a certain canonical basis of holomorphic 1-forms on the RS.
We call this the ``canonical frame" for the SW data and adopt it in this paper; it is reviewed in section \ref{sec3}.
The right hand side of the SW curve in \eqref{SWcurve} can always be written as 
\begin{align}\label{curveform}
    c(x;u,v) \doteq r(u,v) \, p(x;u,v)
\end{align}
where $r$ is a rational function of the CB coordinates and $p$ is polynomial in $x$, $u$, and $v$. 
The classification attempt in \cite{Argyres:2005pp, Argyres:2005wx} was (admittedly) incomplete:  it imposed an additional technical assumption that $p$ in \eqref{curveform} be a \emph{monic} polynomial in $x$; \emph{i.e.}, that $r$ can be chosen such that the leading coefficient in $x$ is 1.
Many CB geometries of known SCFTs were not found.
For instance, CB geometries of the $\cN=4$ sYM theories were not found, neither were those of the $\cN=3$ theories discovered since then \cite{Garcia-Etxebarria:2016erx, Aharony:2016kai, Argyres:2019ngz, Kaidi:2022lyo}, nor were  many CB geometries associated to purely $\cN=2$ theories \cite{Martone:2021ixp}.
This last set will be constructed in this paper.

In this series of papers, comprising the current one and two more \cite{Argyres:2022puv,Argyres:2022fwy}, we reexamine this classification of genus 2 SW curves with the aim of making it complete.
We find that there are four distinct ways in which the strategy of \cite{Argyres:2005pp, Argyres:2005wx} must be generalised or modified in order to capture all rank 2 CB geometries and their corresponding SCFTs:
\begin{itemize}
    \item[{\bf 1.}] The ansatz for $c(x;u,v)$ should be generalised and the search optimised.
    \item[{\bf 2.}] A deformation analysis mapping scale invariant solutions to actual SCFTs can be performed.
    \item[{\bf 3.}] Automorphism groups of genus 2 RSs can be utilised to reduce the calculational complexity of the search for the most general solution.
    \item[{\bf 4.}] A family of curves describing degenerate genus 2 RSs must be included.
\end{itemize}
We now briefly explain the reason for each  of these generalisations.

\paragraph{1. Generalising and optimising the strategy.}

Even with the monic assumption for $p$ in \eqref{curveform}, the special K\"ahler (SK) conditions on the curve ansatz for $c(x; u,v)$ considered in \cite{Argyres:2005pp, Argyres:2005wx} resulted in a complicated system of 33 5th order polynomial equations in 21 unknowns.
One complicating fact was that the scaling dimensions, $(\D_u,\D_v)$, of the CB coordinates were among the unknowns, and for unspecified values of the $\D_i$ there are many possible terms in $p$ which are permitted by scaling.
Since then, the set of allowed pairs of rank-2 scaling dimensions has been independently determined \cite{Argyres:2018zay, Caorsi:2018ahl, Argyres:2018wxu, Cecotti22}, and so can be used as an input to the SK conditions, substantially simplifying them.
The allowed pairs of scaling dimensions are listed in table \ref{tab:CBSDr2}, below.
Furthermore, as we review in section \ref{sec2}, an invariant of scale-invariant CB geometries called the \emph{characteristic dimension} \cite{Cecotti:2021ouq} can also be leveraged to make the scan considerably more efficient.

Even with these simplifications, the SK conditions are algebraically complicated.
But we notice that, even though the monic ansatz used in \cite{Argyres:2005pp, Argyres:2005wx} allowed non-polynomial coefficients of the curve, all the solutions%
\footnote{Note that \cite{Argyres:2005pp, Argyres:2005wx} mistakenly reported some solutions with non-rational (fractional power) prefactors.
These solutions are in fact unphysical since they lead to ill-defined phases of the central charge of the low energy supersymmetry algebra on the CB.}
found were in fact polynomial in $u$, $v$.
Furthermore, there is some (admittedly thin) evidence that non-polynomial $c(x;u,v)$ solutions seem to be associated with geometries which are isotrivial or whose SW curves have extra automorphisms.
These cases may be more amenable to a different calculational strategy, described in paragraph {\bf 3} below.
For these reasons, we specialize in this paper to the case, which we call the \emph{polynomial ansatz}, where 
\begin{align}\label{poly ansatz}
    c(x,u,v) \ \, \text{is assumed polynomial in $x$, $u$, and $v$.}
\end{align} 
With this ansatz the SK conditions further simplify permitting a complete solution, giving the 14 geometries recorded in table \ref{tab:knownSW}.

\paragraph{2. From scale invariance to SCFTs.}

It is well known that each scale invariant CB geometry can be interpreted as the CB of multiple SCFTs \cite{Seiberg:1994rs, Seiberg:1994aj}. 
What is much less clear is how to go about classifying all possible consistent interpretations of each given solution. 
This was done at rank 1 by classifying all possible inequivalent relevant (in the renormalization group sense) SK deformations of a given scale invariant CB geometry \cite{Argyres:2015ffa, Argyres:2015gha, Argyres:2016xua, Argyres:2016xmc}.
This was a highly non-trivial task which brought many surprises and relied heavily on the zero-dimensional nature of the singular locus at rank 1, making it difficult to directly generalise to higher ranks. 
By combining old insights from the rank 1 analysis with new ones involving the stratification of the singular locus \cite{Martone:2020nsy, Argyres:2020wmq}, we have been able to find an indirect and computationally simple way to carry out the deformation analysis at rank 2.
Using this method we are able to map each scale-invariant rank 2 CB geometry to the set of all possible consistent SCFT interpretations for all the geometries we find here.
While we report some of these results in section \ref{sec:6} of this paper, this analysis will be the focus of the second paper in this series \cite{Argyres:2022puv}.
This analysis uncovers three new rank 2 SCFTs which were not known before and sharpens our understanding of three more.

\paragraph{3. Genus 2 RSs with extra automorphisms.}

All genus 2 RSs are hyperelliptic, meaning they all possess a $\Z_2$ group of automorphisms generated by the map $(x,y) \mapsto (x,-y)$ interchanging the two sheets of \eqref{SWcurve}.
Certain special sets of genus 2 RSs have larger (though always finite) automorphism groups.
For these special symmetrical RSs it is calculationally advantageous to fix the reparameterization freedom of the SW curve and 1-form basis by choosing a canonical form for the curve while leaving the 1-form basis arbitrary.
We call this choice of gauge fixing the ``automorphism frame".
The automorphism frame is the opposite of the canonical frame used in this paper (which fixes the 1-form basis to a canonical form).
The subsets of genus-2 RSs with additional automorphisms are described in the third paper of the series \cite{Argyres:2022fwy}, where we also illustrate the use of the automorphism frame to compute an example of a SW curve which, once transformed into the canonical frame, has non-polynomial $c(x;u,v)$.

\paragraph{4. Degenerate genus 2 RSs.}

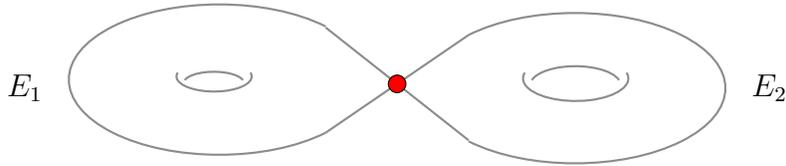
\begin{figure}[ht]
\centering
\begin{tikzpicture}
\draw[rotate=0,thick,draw=black!45] (1.6,-0.6) arc (315:45:2.0cm and 1.0cm);
\node at (-2.4,0) {$E_1$};
\draw[rotate=0,thick,draw=black!45] (0.6,.2) arc (375:165:.5cm and .2cm);
\draw[rotate=0,thick,draw=black!45] (0.5,.1) arc (15:165:.4cm and .15cm);
\draw[rotate=0,thick,draw=black!45] (3.5,0.7) arc (135:-135:2.0cm and 1.0cm);
\node at (7.5,0) {$E_2$};
\draw[rotate=0,thick,draw=black!45] (5.6,.2) arc (375:165:.7cm and .3cm);
\draw[rotate=0,thick,draw=black!45] (5.5,.1) arc (15:165:.6cm and .25cm);
\draw[thick,draw=black!45] (1.6,-0.6) -- (3.5,0.7);
\draw[thick,draw=black!45] (1.6,0.8) -- (3.5,-0.7);
\node[R] at (2.55,0.05) {};
\end{tikzpicture}
\caption{A genus-2 RS pinched to a bouquet of two genus-1 RSs,  $E_1$ and $E_2$.}
\label{fig1}
\end{figure}

The degenerate genus-2 RSs which must be included are those which are ``pinched" to form a bouquet of two genus-1 RSs, as shown in figure \ref{fig1}. 
Though the RS is degenerate, its Jacobian variety is not. 
Instead, the Jacobian variety for this pinched RS simply splits into the product of two rank-1 varieties (complex tori), meaning that there is a coordinate basis in which the $2\times2$ symmetric complex matrix $\t_{ij}$ describing the modulus of the Jacobian variety is diagonal,
\begin{align}\label{split tau}
    \t = \bpm z_1 & 0 \\ 0 & z_2\epm, \qquad \Im z_1 >0 \quad\text{and}\quad \Im z_2 >0 .
\end{align}
Since it is the Jacobian variety (and not directly the genus-2 RS) which is what determines the geometry of the CB, a family of such ``split" RSs, though degenerate everywhere on the CB, may still describe a regular CB geometry.
A family of everywhere split RSs cannot be described by a family of binary-sextic plane curves \eqref{SWcurve}.
Instead, it requires a different presentation as a pair of elliptic curves. 
We note that the split case always arises when the CB has characteristic dimension not equal to $1$ or $2$; this is explained in section \ref{sec2} below.
The analysis of the split case will be left for future work.

\vspace{5mm}

As will be explained in detail in \cite{Argyres:2022puv}, it is remarkable that every solution that we find does have an interpretation as the CB of at least one known $\cN=2$ SCFT. 
We also highlight that the solutions described here, joined with those with characteristic dimension not equal $1$ or $2$ (which are in the split case) already capture the majority of the rank 2 theories currently known. 
But generalising our search here beyond the polynomial ansatz --- perhaps best addressed by searching in suitable automorphism frames --- could produce new solutions. 
So it is still likely that this line of investigation will ultimately produce new CB geometries corresponding to currently unknown $\cN=2$ SCFTs.

\begin{center}
\rule[1mm]{2cm}{.4pt}\hspace{1cm}$\circ$\hspace{1cm} \rule[1mm]{2cm}{.4pt}
\end{center}

Before starting our main presentation, it is important to highlight the status of the two main physical assumptions that underlie our program:
\vspace{2mm}

\noindent\textit{Absence of complex singularities}. 
There are known examples of $\cN=2$ SCFTs whose CBs have complex singularities \cite{Argyres:2018wxu}. 
These can all be obtained from SCFTs whose CBs have no complex singularities by gauging discrete symmetries.
Thus their existence could be inferred from the classification pursued here. 
But there are examples of rank 2 CB geometries with complex singularities which do not arise by discretely gauging other theories, \emph{e.g.}, three of the orbifold geometries found in \cite{Argyres:2019ngz}.
On the other hand, there is no separate evidence for the existence of $\cN{=}2$ SCFTs with these CBs. 
\vspace{2mm}

\noindent \textit{Principal polarisation}. 
The many $\cN{=}2$ gauge theories which do not have principal Dirac pairing are examples of \emph{relative} field theories \cite{DelZotto:2022ras}.
By maximally refining the charge lattice by adding probe line charges (corresponding to a choice of global structure of the gauge theory) they can be made into genuine field theories with principal pairing  \cite{Gaiotto:2010be, Tachikawa:2013hya, Tachikawa:2013kta, Argyres:2022kon}. 
These lagrangian SCFT cases are thus covered by our classification strategy.
But, already at rank 2, there are also (non-lagrangian) SCFTs whose CB has no principally polarised version \cite{Caorsi:2018zsq, Argyres:2022kon}. 
An example is the $\cN{=}3$ SCFT associated to the complex reflection group $G_8$ whose M-theory realisation was presented in \cite{Garcia-Etxebarria:2016erx, Kaidi:2022lyo}. 

\vspace{2mm}

The rest of this paper is organised as follows. 
Section \ref{sec2} briefly reviews the basic properties of rank 2 CB geometries and recalls the definition of the characteristic dimension of an $\cN{=}2$ SCFT. 
Section \ref{sec3} presents the binary-sextic plane curve form \eqref{SWcrv} for the SW curve, and introduces the polynomial ansatz. 
Section \ref{sec4} discusses the conditions on the SW data coming from $\cN{=}2$ supersymmetry, scale-invariance, and unitarity, and introduces the SK integrability conditions the SW curves must satisfy;  this is largely a summary of a discussion given in \cite{Argyres:2005pp}. 
We show how we organise the search for solutions to the integrability conditions and present one solution in detail. 
The other new solutions that we find here --- those highlighted in green in table \ref{tab:knownSW} --- are presented in section \ref{sec:5}.
For completeness we also present the solutions found in \cite{Argyres:2005pp, Argyres:2005wx} in appendix \ref{appA}. 
Finally, in section \ref{sec:6}, we summarise how the solutions we find fit in the bigger picture of rank-2 SCFTs. 
A much more detailed analysis of this last point will be reported in \cite{Argyres:2022puv}.

\section{Review of rank 2 Coulomb branch geometries}\label{sec2}

Before delving into the details of solving the constraints on CB geometries stemming from $\cN{=}2$ superconformal invariance, it is useful to review some general results on the structure of CBs of $\cN{=}2$ SCFTs.

Recall first that scale invariant CB geometries enjoy a holomorphic $\C^*$ symmetry action coming from the combination of the dilatation and $\U(1)_R$ symmetries spontaneously broken on the CB. 
We will make the assumption that the CB has no complex singularities and thus is simply $\C^r$ as a complex space. 
Here $r$ is the rank of the CB, which is simply its complex dimension.
In this case we can choose complex coordinates, $(u_1,...,u_r) \in \C^r$, unique up to normalization, which diagonalize the $\C^*$ action,
\begin{align}\label{Cstar}
\C^* : (u_1,...,u_r) &\mapsto (t^{\D_1} u_1,..., t^{\D_r} u_r) , &
t &\in\C^* .
\end{align}
We refer to $(\D_1,...,\D_r)$ as the CB scaling dimensions; they will play a key role in all that follows.

While it is unknown whether at each rank there are finitely many scale invariant special geometries, it is by now established that at any given rank the allowed set of scaling dimensions is finite and rational \cite{Argyres:2018zay, Caorsi:2018ahl, Argyres:2018urp}.
At rank 2 name the CB scaling coordinates 
\begin{align}\label{uvdef}
    (u_1,u_2) \doteq (u,v) \in \C^2, \qquad \text{with} \quad \D_u \le \D_v .
\end{align}
In total at rank 2 there are 79 allowed pairs $(\D_u,\D_v)$ of scaling dimensions, given by a refinement of an argument in \cite{Caorsi:2018ahl, Cecotti22}.
They are listed in table \ref{tab:CBSDr2}. 
Our search for rank 2 CB geometries will be organised on a case by case basis using these scaling dimension pairs as input. 

\begin{table}[!t]
\begin{center}
\renewcommand{\arraystretch}{1.2}
\hspace*{-0.1cm}\begin{tabular}{c|c}
\multicolumn{2}{c}{\textsc{Genuinely rank 2 pairs}}\\
\hline\hline
$n$& $\{\D_u,\D_v\}$
\\\hline\hline
5
&$\{\frac{4}{3},\frac53\}\ \{\frac54,\frac32\}\ \{\frac32,\frac52\}\ \{\frac54,3\}\ \{\frac53,3\}\ \{\frac52,3\}\ \{3,5\}\ \{\frac54,4\}\ \{\frac53,4\}\ \{\frac52,4\}\ \{4,5\}$\\
\hline
5,8
&$\{\frac54,8\}\ \{\frac53,8\}\ \{\frac52,8\}\ \{5,8\}$\\
\hline
8
&$\{\frac65,\frac85\}\ \{\frac43,\frac83\}\  \{\frac87,4\}\ \{\frac85,4\}\ \{\frac83,4\}\ \{4,8\}\ \{\frac87,6\}\ \{\frac85,6\}\ \{\frac83,6\}\ \{6,8\}$\\
\hline
8,10
&$\{\frac87,\frac{10}7\}\ \{\frac83,\frac{10}3\}\ \{\frac{10}9,8\}\ \{\frac{10}7,8\}\ \{\frac{10}3,8\}\ \{8,10\}$\\
\hline
8,12
&$\{\frac87,\frac{12}7\}\ \{\frac85,\frac{12}5\}\ \{\frac{12}{11},8\}\ \{\frac{12}7,8\}\ \{\frac{12}5,8\}\ \{\frac87,12\}\ \{\frac85,12\}\ \{\frac83,12\}\  \{8,12\}$\\
\hline
10
&$\{\frac{10}{9},\frac43\}\ \{\frac43,\frac{10}3\}\ \{\frac{10}9,4\}\ \{\frac{10}7,4\}\ \{\frac{10}3,4\}\ \{4,10\}$\\
\hline 
12
&$\{\frac65,\frac{12}5\}\ \{\frac{12}{11},6\}\ \{\frac{12}7,6\}\ \{\frac{12}5,6\}\ \{6,12\}$\\
\hline\hline
\multicolumn{2}{c}{\textsc{Non-Genuinely rank 2 pairs}}\\
\hline\hline
\multicolumn{2}{c}{$\{\D_u,\D_v\}$}\\
\hline\hline
\multicolumn{2}{c}{$\{\frac65,\frac65\}\ \{\frac65,\frac43\}\ \{\frac65,\frac32\}\ \{\frac65,2\}\ \{\frac65,3\}\ \{\frac65,4\}\ \{\frac65,6\}\ \{\frac43,\frac43\}\ \{\frac43,\frac32\}\ \{\frac43,2\}$}\\
\multicolumn{2}{c}{$\{\frac43,3\}\ \{\frac43,4\}\ \{\frac43,6\}\ \{\frac32,\frac32\}\ \{\frac32,2\}\ \{\frac32,3\}\ \{\frac32,4\}\ \{\frac32,6\}\ \{2,2\}\ \{2,3\}$}\\
\multicolumn{2}{c}{$\{2,4\}\ \{2,6\}\ \{3,3\}\ \{3,4\}\ \{3,6\}\ \{4,4\}\ \{4,6\}\ \{6,6\}$}\\
\hline\hline
\end{tabular}
\end{center}
\caption{All 
79 allowed pairs of scaling dimensions at rank 2. 
The genuinely rank 2 pairs are those which include at least one entry which is not allowed at rank 1. 
For convenience we divide the genuinely rank 2 pairs into sets labeled by the numerators, $n$, of the genuinely rank 2 scaling dimensions.
}
\label{tab:CBSDr2}
\end{table}%

In fact, the restriction that the generic RS fibre of the SW curve is non-degenerate will allow us to only have to work with a subset of all allowed pairs. 
To explain this, we first recall the basics of the special K\"ahler structure of the CB.
The low energy effective theory on a generic point $u$ of a rank $r$ CB is an $\cN{=}2$ $\U(1)^r$ gauge theory defined only in terms of its $u$-dependent matrix of holomorphic gauge couplings, $\t_{ij}(u)$. 
This structure is encoded in the special geometry of the CB which is a $\C^*$-isoinvariant holomorphic integrable system, \emph{i.e.}, a holomorphic fibration over the CB, $\pi: X\to \C^r$, such that the total space $X$ is holomorphic-symplectic with lagrangian fibres \cite{Seiberg:1994rs, Seiberg:1994aj, Donagi:1995cf} (see also \cite{Freed:1997dp}).
The smooth fibres are polarized abelian varieties of dimension $r$ whose period matrices $\t_{ij}(u)$ are equal to the effective couplings.

The special geometry is \emph{isotrivial} iff all smooth fibres are isomorphic (as polarized abelian varieties) to a fixed abelian variety $A$, in other words if the matrix of effective couplings $\tau_{ij}$ is actually constant over the entire CB. 
The special geometry is \emph{diagonal} iff, in addition, the matrix of effective couplings is proportional to the identity,
\beq
\tau_{ij}=\z\, \I_{r\times r} ,
\eeq
where $\z$ is a root of unity.
This means that for a diagonal CB, the fixed abelian variety is the product of $r$ rank 1 abelian varieties, $A\cong E_\z \times E_\z \times \cdots \times E_\z$, where $E_\z$ is the elliptic curve with period $\tau\equiv \zeta$. 
At rank 2 this corresponds to having $A$ being the Jacobian torus of the aforementioned ``split" genus 2 RS, a case which will not be considered here. 

Diagonal geometries are very special --- though less rare than one might think --- and they have a large automorphism group which is only compatible with special values of the CB scaling dimensions. 
The key invariant which identifies the diagonal geometries is the \emph{characteristic dimension}, $\varkappa$ \cite{Cecotti:2021ouq}, defined by
\begin{align}
    \varkappa \doteq \frac1{\{\l^{-1}\}} ,
\end{align}
where $\l \doteq \gcd(\D_1,\ldots,\D_r)$ and $\{\l^{-1}\} \doteq \l^{-1} \mod 1$ with $0<\{\l^{-1}\}\leq 1$.  
It is shown in \cite{Cecotti:2021ouq} that $\varkappa$ can take only one of eight values,
\begin{align}
    \varkappa \in \left\{1,\frac 65,\frac 43,\frac 32,2,3,4,6 \right\},
\end{align}
and that if $\varkappa\notin\{1,2\}$ then the CB geometry is necessarily diagonal.%
\footnote{Note that there are several isotrivial special geometries which are not diagonal; these necessarily have $\varkappa=1$ or 2. 
$\cN{=}4$ SCFTs give examples of non-diagonal isotrivial CB geometries.} 

This last property considerably reduces the number of allowed pairs of CB scaling dimensions at rank 2 that we need to consider, since, by what we have discussed so far, only those corresponding to $\varkappa=1$ or 2 can be described by generically regular genus 2 SW curves. 
It is a straightforward exercise to check that out of the 79 allowed pairs of scaling dimensions at rank 2, only 47 have $\varkappa=1$ or $2$.
These 47 pairs are listed in table \ref{tab:CBSDr2kappa}. 
This means that CB geometries for the 32 scaling dimension pairs not in table \ref{tab:CBSDr2kappa} are necessarily diagonal and need to be analysed using families of split RSs instead of the families of regular genus 2 RSs considered here.

\begin{table}[!t]
\begin{center}
\renewcommand{\arraystretch}{1.2}
\hspace*{-0.1cm}\begin{tabular}{c|c}
\multicolumn{2}{c}{\textsc{CB dimensions with $\varkappa =1$ or $2$}}\\
\hline\hline
& $\{\D_u,\D_v\}$
\\\hline\hline
\multirow{3}{*}{Polynomial ansatz}&$\{\frac{10}{9},\frac43\}\ \{\frac87,\frac{12}7\}\ \{\frac65,\frac85\}\  \{\frac54,\frac32\}\ \{\frac{4}{3},\frac53\}\ \{\frac{4}{3},2\}\ \{\frac32,2\}\ \{\frac32,\frac52\}\  \{\frac53,3\}\  $\\
&$\{2,2\}\ \{2,3\}\ \{2,4\}\ \{\frac52,3\}\ \{\frac52,4\}\ \{\frac83,\frac{10}3\}\ \{\frac83,6\}\  \{3,4\}\  \{3,5\}\ $\\
&$\{\frac{10}3,4\}\ \{\frac{10}3,8\}\ \{4,5\}\  \{4,6\}\  \{4,10\}\ \{5,8\}\ \{6,8\}\ \{8,10\}  $\\
\hline
\multirow{3}{*}{Non-polynomial}&$ \{\frac{10}9,4\}\  \{\frac{10}9,8\}\  \{\frac87,6\}\ \{\frac65,\frac43\}\  \{\frac65,2\}\ \{\frac65,4\}\  \{\frac54,3\}\ \{\frac54,4\}\ \{\frac54,8\}\ $\\
&$\{\frac43,\frac32\}\ \{\frac43,3\}\ \{\frac43,\frac{10}3\}\ \{\frac43,6\}\  \{\frac{10}7,4\}\  \{\frac{10}7,8\}\  \{\frac32,4\}\   \{\frac85,6\}\  $\\
&$\{\frac53,4\}\   \{\frac53,8\}\  \{2,6\}\  \{\frac52,8\}\  $\\
\hline\hline
\end{tabular}
\end{center}
\caption{The 47 allowed pairs of scaling dimensions at rank 2 with characteristic dimension equal to one or two. These are further divided into the 26 pairs for which a SW curve with polynomial coefficients is allowed, and the 21 pairs for which it isn't.
}
\label{tab:CBSDr2kappa}
\end{table}%

\section{Ansatz for the Seiberg-Witten curve}
\label{sec3}

We have already simplified our problem remarkably while still maintaining full generality. 
To make further progress we have to discuss in more detail how to algebraically encode the CB geometry.

At rank 2 all principally polarized abelian varieties arise as the Jacobian variety of a genus-2 Riemann surface or of a singular limit of such a surface. 
Furthermore, all non-singular genus-2 Riemann surfaces can be written as a suitable projectivization%
\footnote{The suitable projectivization is $y^2 = c(x,w)$ where $c$ is homogeneous of degree 6 in $x$ and $w$ and $[w:x:y] \in \P^2_{(1,1,3)}$ are homogeneous coordinates of a weighted projective space.}
of binary sextic plane curves as in \eqref{SWcurve},
\begin{align}\label{SWcrv}
y^2 &= c(x,u,v) \doteq \sum_{a=0}^6 c_a x^a ,
\qquad c_6\ \text{or}\ c_5 \neq 0 .
\end{align}
The curve, by itself, does not completely determine the CB effective action: the central charge $Z$ of the low energy supersymmetry algebra on the CB must also be specified.
The central charge is a function of the electric and magnetic charge sector, and its norm gives the BPS mass in that sector.
In particular, the derivatives of $Z$ with respect to the CB coordinates are given by periods of a basis of holomorphic 1-forms on the RS.

The curve and 1-form basis data is highly redundant, as there is a $\GL(2,\C)$ group of reparametrisations of $x$ and $y$ which leave the algebraic form of the curve \eqref{SWcrv} invariant, while changing the 1-form basis.
This reparametrisation freedom can be completely fixed by bringing the 1-form basis to the canonical form
\beq\label{forms}
\w_u \doteq \frac{xdx}y, \qquad \w_v \doteq \frac{dx} {y}.
\eeq
With this basis, the central charge satisfies
\beq\label{CC}
\del_u Z=\oint \w_u\quad{\rm and}\quad \del_v Z=\oint \w_v,
\eeq
where the homology class of the integration cycles encodes the electric-magnetic charge sector.

We call this way of fixing the curve reparametrisation freedom the \emph{canonical frame}.
The $(x,y)$ reparametrisation freedom and different ways of fixing it are discussed in some detail in \cite{Argyres:2022fwy}, where it is also shown that in canonical frame the $c_a$ coefficients of the curve are meromorphic in $(u,v)$. 
The choice of canonical frame still leaves unfixed the reparameterization invariance of the CB coordinates $(u,v)$.
Since we are using global coordinates, this freedom --- what was called the ``holomorphic reparametrisation invariance" in \cite{Argyres:2005pp} --- is very constrained.
In particular, once $u$ and $v$ are chosen as coordinates with definite scaling dimensions, the only freedom left is overall constant rescalings, 
\begin{align}\label{CBreparam}
    (u,v) &\mapsto (\xi_u u, \xi_v v)&
    &\text{with}& 
    \xi_i&\in\C^*&
    &\text{if}& 
    \D_u &\nmid \D_v.
\end{align}
We will use this CB reparametrisation freedom to fix two non-zero coefficients of the curve solutions we find to convenient values. 
In case $\D_u |\D_v$, there is a slightly larger invariance under redefinitions $(u,v) \mapsto (\xi_u u, \, \xi_v v{+}\eta u^p)$ with $\xi_i\in\C^*$ and $\eta\in\C$ and where $p \doteq \D_v/\D_u$, which can be used to set 3 curve coefficients to arbitrary values.
Finally, when $\D_u=\D_v$, general linear redefinitions of $u$ and $v$ can be used to set 4 curve coefficients to arbitrary values.

Because of scale invariance, \eqref{SWcrv} must be homogeneous with respect to the $\C^*$ action \eqref{Cstar}.
Assigning scaling dimensions $\D_x$ and $\D_y$ to the curve coordinates and $\D_{c_a}$ to the curve coefficients, we therefore have
\beq\label{scaleca}
\D_{c_a}=2\D_y-a \D_x,\quad a=0,...,6.
\eeq
Since the scaling dimension of $Z$ is by definition one --- it gives the masses of BPS states in the theory --- we can use \eqref{CC} to recast the scaling dimension of $x$ and $y$ in terms of those of $u$ and $v$,
\beq\label{scalexy}
    \D_x=\D_v-\D_u
    \quad \text{and} \quad 
    \D_y=2\D_v-\D_u-1 .
\eeq
Therefore, using \eqref{scaleca},
\beq\label{SCal}
\D_{c_a}=(4-a)\D_v-(2-a)\D_u-2,\quad a=0,...,6.
\eeq

We now impose an additional, technical, simplifying assumption, which we call the \emph{polynomial ansatz} for the SW curve.
This is the assumption that the $c_a$ are holomorphic functions on the CB.
It follows from the constraints from scale invariance \eqref{SCal}, that they are in fact polynomial in $u$ and $v$. 
As discussed in the introduction, this assumption is not well motivated from a physical standpoint, since there are CB geometries for which the curve coefficients have poles on the CB.
We adopt it here merely for calculational convenience, and understanding how to lift it while preserving a finite algorithmic search for solutions is the main technical challenge faced by our approach to systematically classifying all rank 2 scale invariant CB geometries.
It is interesting to note that even though the monic ansatz used in \cite{Argyres:2005pp, Argyres:2005wx} covered cases beyond the polynomial ansatz, all solutions found there turned out to be polynomial and are thus reproduced by our analysis here.
An alternate approach, using what we call \emph{automorphism frames} as opposed to the canonical frame used here, is explored in \cite{Argyres:2022fwy}.

To appreciate how constraining the polynomial ansatz is, observe that there are many pairs of scaling dimensions with $\varkappa\in\{1,2\}$ for which no polynomial ansatz is even possible, since scale invariance does not allow any non-zero $c_5$ or $c_6$ coefficient. 
For example, take $\{\D_u,\D_v\}=\{\frac{10}9,4\}$ from table \ref{tab:CBSDr2kappa}. 
Using \eqref{SCal} we find
\begin{align}\label{10940}
\{\D_u,\D_v\}&=\{\tfrac{10}9,4\} &
&\Rightarrow& &
\begin{array}{c|c|c|c|c|c|c|c}
   &\ c_0\ \,&\ c_1\ \,&\ c_2\ \,&\ c_3\ \,&\ c_4\ \,&\ c_5\ \,&\ c_6\ \, \\
\hline
\D\ \, &\frac{106}9&\frac{80}9&6&\frac{28}9&\frac29&-\frac{8}3&-\frac{50}9
\end{array}
\end{align}
from which we can infer that the most general polynomial ansatz for this case is
\begin{align}\label{10941}
y^2=\l_1 u^8 x+\l_2 u^7 v 
\end{align}
which is linear in $x$ and thus not a genus 2 RS. Going through the list of CB scaling dimension pairs and seeking those which admit non-zero $c_5$ or $c_6$ coefficient allows to divide the scaling dimensions as shown in table \ref{tab:CBSDr2kappa}.
This further restricts our search to only 26 pairs of scaling dimensions. 
In order to proceed in our analysis we have to finally discuss the constraints on the curve following from superconformal invariance to which we turn next.

\section{Special K\"ahler integrability condition}
\label{sec4}

The existence of the symplectic form on the bundle of Jacobians over the CB is equivalent to the existence of a SW 1-form, $\L$, a meromorphic 1-form on the genus-2 RS whose periods give the central charge,
\begin{align}
    Z = \oint \L .
\end{align}
The central charge conditions \eqref{CC} then imply that $\L$ satisfies the differential equations
\begin{align}\label{SW1f}
\del_i \L &= \w_i +df_i, & i &\in \{u,v\},
\end{align}
where $\del_i \doteq \del/\del i$, $\w_i$ is the canonical basis \eqref{forms} of holomorphic 1-forms on the RS, $f_i$ are undetermined meromorphic functions on the RS, and $d$ is the exterior derivative on the RS.  
$\L$, $\w_i$, and $f_i$ all vary holomorphically over the CB.  
The existence of a $\L$ satisfying \eqref{SW1f} ensures that the CB has a special K\"ahler (SK) structure.

A necessary condition for the existence of a SW 1-form satisfying \eqref{SW1f} is the \emph{SK integrability condition}
\begin{align}\label{SWint}
\del_u\w_v - \del_v \w_u = d g,
\end{align}
where $g$ is a meromorphic function on the RS. 
In the case of a scale-invariant CB, the integrability condition is also a sufficient condition for the existence of the SW 1-form, since it is uniquely determined by scale invariance to be the holomorphic 1-form
\begin{align}\label{}
\L = (1+\D_x)^{-1} (\D_u u\, \w_u + \D_v v\, \w_v) ,
\end{align}
if \eqref{SWint} is satisfied.

As argued in \cite{Argyres:2005pp}, the most general form of the meromorphic function $g$ that can appear on the right side of \eqref{SWint} is
\begin{align}\label{bfun}
g = \frac{b(x,u,v)}{y} \doteq \sum_{s=0}^3 b_s \frac{x^s}y .
\end{align}
where $b_s$ are meromorphic functions of $u$ and $v$.
Using the canonical form \eqref{forms} of the 1-form basis, \eqref{SWint} can be recast as
\beq\label{integr}
\del_u\left(\frac 1 y\right)-\del_v\left(\frac x y\right)-\del_x\left(\frac b y\right)=0 ,
\eeq
where $b$ is the meromorphic function of $u$ and $v$ which is at most cubic in $x$ defined in \eqref{bfun}. 
Using the explicit form \eqref{SWcrv} for $y$ gives
%
\beq\label{intEq}
b\frac{\del c}{\del x}-\frac{\del c}{\del u}{-}x\frac{\del c}{\del v}-2 c\frac{\del b}{\del x}=0 ,
\eeq
\noindent where $c\equiv c(x,u,v)$ is defined in \eqref{SWcrv} and $b\equiv b(x,u,v)$ in \eqref{bfun}. 
This differential equation is the key equation that we need to solve.

Solving \eqref{intEq} is a daunting problem for general meromorphic $b$ and $c$, but simplifies dramatically for polynomial $c$.
Our strategy is to:
\begin{itemize}
    \item[1.] go through the list of pairs of scaling dimensions in the first row of table \ref{tab:CBSDr2kappa},
    \item[2.] compute for each the corresponding polynomial ansatz for $c$, and
    \item[3.] look for solution for \eqref{intEq}.
\end{itemize}
In this case \eqref{intEq} reduces to a set of eight algebraic equations in the $b_s(u,v)$, $s=0,1,2,3$, and the constants $\l_i$ parametrising the terms compatible with scale invariance in $c(x,u,v)$ as in \eqref{10940}--\eqref{10941}. 
Notice that naively \eqref{intEq} is a degree 8 polynomial in $x$, but the $x^8$ coefficient vanishes identically. 
Finally we also have to check that the solution we get is a non-singular genus 2 curve for generic $(u,v)$.
That is, we check that either $c_5$ or $c_6$ is not vanishing at the generic point of the CB, and also that the $x$ discriminant of $c(x;u,v)$ does not vanish identically. 
Finally we can use the CB reparametrisation invariance \eqref{CBreparam} to fix two non-zero coefficients to any values which makes the curve look particularly simple.

All the non-singular solutions of \eqref{intEq} we find with polynomial ansatz for $c$ have $b \equiv 0$.  
We do not know why this had to be the case; there are many solutions with $b\neq0$ with singular polynomial $c$ as well as regular non-polynomial $c$.
Given that $b\equiv 0$, the general polynomial solution of \eqref{intEq} is easily seen to be of the form
\begin{align}
    c(x;,u,v) = P(x, ux-v)
\end{align}
where $P$ is a polynomial in its two arguments.

\subsection{Detailed example of a new solution: $\{\D_u,\D_v\}=\{4,5\}$}
\label{sec:45}

As an illustration of how all this works, we carry out the analysis in a simple case: $\{\D_u,\D_v\}=\{4,5\}$. 
All the remaining cases are reported in the next section or in appendix \ref{appA}.

First of all,
\beq
\varkappa(\{4,5\})=1,
\eeq
so for this pair of scaling dimensions the generic fibre is non-singular and the analysis above applies. 
Now using \eqref{SCal} we can readily compute the scaling dimension of the $c_a$'s,
\begin{equation}
\{4,5\}:\quad
\begin{array}{c|c|c|c|c|c|c|c}
 & c_0 & c_1 & c_2 & c_3 & c_4 & c_5 & c_6 \\\hline
\D& 10 & 9 & 8 & 7 & 6 & 5 & 4 \\
\end{array}
\end{equation}
from which we obtain the polynomial ansatz
\begin{equation}\label{crv45}
y^2=v^2 \l_1+u v x \l_2+u^2 x^2 \l_3+v x^5 \l_4+u x^6 \l_5 .
\end{equation}
Since both an $x^6$ and an $x^5$ term are allowed, we don't know which of $\l_4$ or $\l_5$ is non-vanishing, so we do not use the CB reparametrizations \eqref{CBreparam} to set either to 1.
The SK integrability condition \eqref{intEq} leads to the system of equations in the constants $\l_a$ and meromorphic functions $b_s(u,v)$,
\begin{align}
\cO(x^0):&& 0&= -2 v^2 b_1 \l_1+u v b_0 \l_2 \nn\\
\cO(x^1):&& 0&= -2 v \l_1-4 v^2 b_2 \l_1-v \l_2-u v b_1 \l_2+2 u^2 b_0 \l_3 \nn\\
\cO(x^2):&& 0&= -6 v^2 b_3 \l_1-u \l_2-3 u v b_2 \l_2-2 u \l_3 \nn\\
\cO(x^3):&& 0&= -5 u v b_3 \l_2-2 u^2 b_2 \l_3 \\
\cO(x^4):&& 0&= -4 u^2 b_3 \l_3+5 v b_0 \l_4 \nn\\
\cO(x^5):&& 0&= 3 v b_1 \l_4+6 u b_0 \l_5 \nn\\
\cO(x^6):&& 0&= -\l_4+v b_2 \l_4-\l_5+4 u b_1 \l_5 \nn\\
\cO(x^7):&& 0&= -v b_3 \l_4+2 u b_2 \l_5 \nn
\end{align}
which has many solutions.
All but one of the solutions have either $\l_4=\l_5=0$ --- which not a genus 2 curve --- or $\l_1=\l_2=\l_3=0$ in which case \eqref{crv45} degenerates to $y^2=x^5(v\, \l_4+ u\, \l_5\, x)$
which has five coincident roots at $y=0$ and should be discarded. 
The only remaining solution is
\beq
b_0=b_1=b_2=b_3= 0\quad{\rm and}\quad 
\l_1= -\frac{\l_2}{2},\ 
\l_3= -\frac{\l_2}{2},\ 
\l_5= -\l_4,
\eeq
which, upon setting $\l_4=-1$ and $\l_2=-2$ using the CB reparametrisations \eqref{CBreparam}, gives rise to the curve
\beq\label{45sol}
y^2=
(u\, x-v) (x^5 + u\, x-v) .
\eeq
This is indeed non-singular for generic $(u,v)$ and thus an allowed solution. 
This solution we find is new! 
Indeed, the right side is not a monic polynomial in $x$, which explains why this solution does not appear in the list of consistent scale invariant rank 2 geometries found in \cite{Argyres:2005pp, Argyres:2005wx}.

\subsection{Physical interpretation of the resulting CB geometry}
\label{strat}

Once supplied with a curve \eqref{45sol} in canonical frame, a great deal of information about the physics at the singular locus $\cV$ of the CB can be easily inferred using the results of \cite{Martone:2020nsy, Argyres:2020wmq}.
In particular, the singular locus corresponds to the 1-dimensional subvarieties, ${\mathfrak S}_a$, of the CB along which the curve degenerates.
By scaling, there are only 3 possible types of singular subvarieties,
\begin{align}
    {\mathfrak S}_u & \doteq \{ u=0\},&
    {\mathfrak S}_v & \doteq \{ v=0\},&
    {\mathfrak S}_\text{knot} & \doteq \{ u^q - t v^p=0\}, \quad t\in\C^*,
\end{align}
where $p\doteq \D_u/\gcd(\D_u,\D_v)$ and $q\doteq \D_v/\gcd(\D_u,\D_v)$.
The zeros of the discriminant with respect to $x$ of $c(x;u,v)$ determines the occurrence of each type of ${\mathfrak S}_a$.
The multiplicity of each zero%
\footnote{More correctly, the monodromy of a canonical homology basis of the curve around the singular subvariety.} 
determines the Kodaira type of ${\mathfrak S}_a$.
These facts are reviewed, explained, and analyzed in detail in \cite{Argyres:2022puv}.

The Kodaira types are recorded in table \ref{tab:knownSW} for all polynomial ansatz solutions.
They are physically important because to each ${\mathfrak S}_a$ subvariety is associated a rank 1 SCFT (or infrared-free gauge theory) describing the degrees of freedom which become massless at the singular locus.  
The Kodaira type constrains what these possible rank 1 SCFTs can be to a finite set. 
This information will be used in \cite{Argyres:2022puv} to map each scale invariant CB geometry found here to SCFT data, including flavor symmetry algebra, central charges, Higgs branch dimensions, \emph{etc.}

In our $\{\D_u,\D_v\}=\{4,5\}$ example, this determination goes as follows.
The discriminant of the right hand side of the curve solution \eqref{45sol} is
\begin{align}
    D_x \doteq {\rm Disc}_x \left[(u\, x-v) (x^5 + u\, x-v)\right]
    = v^{10} (256 u^5 + 3125 v^4).
\end{align}
This shows there is no ${\mathfrak S}_u$ singularity, a ${\mathfrak S}_v$ singularity of multiplicity 10, and a ${\mathfrak S}_\text{knot}$ singularity of multiplicity 1.
The only Kodaira type giving multiplicity 1 is $I_1$, while both types $I_{10}$ an $II^*$ give multiplicity 10.
A closer examination of the homology monodromy of the curve around the ${\mathfrak S}_v$ locus shows that it is of type $I_{10}$.
This data is recorded in the ${\mathfrak S}_a$ columns in table \ref{tab:knownSW}.

\section{New scale invariant solutions}\label{sec:5}

In table \ref{tab:CBSDr2kappa} we listed all the pairs of scaling dimensions with characteristic dimension $\varkappa=1$ or 2. 
Of those, half of them do not admit any polynomial ansatz of the SW curve in the form \eqref{SWcrv}; they are listed in the ``non-polynomial" row in the table. 
Then, following the procedure described in the previous section, we found all solutions of the integrability equation \eqref{intEq} for polynomial $c(x;u,v)$. 
Here we describe the new solutions we find (in addition to the $\{4,5\}$ solution described in the previous section).
The ones which were already listed in \cite{Argyres:2005pp, Argyres:2005wx} are reported in appendix \ref{appA}. 
All these solutions are summarized in table \ref{tab:knownSW}.
Remarkably:
\begin{itemize}
    \item Nearly half of the scaling dimension analysed, while admitting a polynomial ansatz for $c(x;u,v)$, do not admit any solution for the integrability equation.  We list these pairs of scaling dimensions in table \ref{tab:NoSols}.
    \item Every solution we find does appear to have an interpretation as a CB of rank-2 SCFTs.  In fact, as happens in rank 1, many of them have multiple SCFT interpretations depending on the chosen deformation pattern of the rank 1 theories supported on their complex co-dimension one singular locus. This will be discussed in detail in \cite{Argyres:2022puv}.
\end{itemize}
That the integrability equation is so constraining, allowing for so few solutions, and that all of these solutions do have a physical interpretation, are remarkable facts, supporting the approach of classifying SCFTs by studying their CB geometries. 

We now briefly record the new solutions we find.

\begin{table}[!t]
\begin{center}
\renewcommand{\arraystretch}{1.2}
\hspace*{-0.1cm}\begin{tabular}{c}
\textsc{CBs with $\varkappa=1$ or $2$ admitting polynomial $c$ but no solution for \eqref{intEq}}\\
\hline\hline
 $\{\D_1,\D_2\}$ \\
\hline\hline
$\{\frac{10}{9},\frac43\}\  \{\frac{4}{3},2\}\ \{\frac32,2\}\   \{\frac53,3\}\ \{\frac52,3\}\ \{\frac52,4\}\ \{\frac83,\frac{10}3\}\ \{\frac83,6\}\  \{\frac{10}3,4\}\ \{\frac{10}3,8\}\ \{5,8\}\ \{8,10\}   $\\
\hline\hline
\end{tabular}
\end{center}
\caption{The 12 out of 26 allowed pairs of scaling dimensions at rank 2 with $\varkappa=1$ or 2 which admit a polynomial ansatz for $c(x;u,v)$, but for which there are no solutions of the integrability equation.
}
\label{tab:NoSols}
\end{table}%

\subsection{$\{3,4\}$}\label{sec:34}

By scaling, the dimensions of the curve coefficients are
\begin{equation}
\{3,4\}:\quad
\begin{array}{c|c|c|c|c|c|c|c}
 & c_0 & c_1 & c_2 & c_3 & c_4 & c_5 & c_6 \\\hline
\D& 8 & 7 & 6 & 5 & 4 & 3 & 2 \\
\end{array}
\end{equation}
from which follows the most general polynomial ansatz,
\begin{equation}\label{ans34}
y^2=v^2 \l_1+u v x \l_2+u^2 x^2 \l_3+v x^4 \l_4+u x^5 \l_5.
\end{equation}
This curve describes a genus two surface only if $\l_5\neq0$, thus in searching for a solution of \eqref{intEq} we can safely assume a non-zero leading coefficient by fixing $\l_5=1$. 
We then find the solution
\beq
b_0=b_1=b_2=b_3= 0\quad{\rm and}\quad 
\l_1= -\frac{\l_2}{2},\ 
\l_3= -\frac{\l_2}{2},\ 
\l_4= -1
\eeq
which, after using reparametrisation invariance to also fix $\l_2\to -2$, leads to the following curve:
\beq\label{34sol}
y^2=(u\, x-v)(x^4 +u\, x-v) .
\eeq
The discriminant of the RHS is
\beq\label{Dis45}
D_x = -v^8  (27 u^4+ 256 v^3)
\eeq
which shows a multiplicity 1 ${\mathfrak S}_\text{knot}$ singularity and a multiplicity 8 ${\mathfrak S}_v$ singularity.
The knotted component has $I_1$ Kodaira type (the only one compatible with an order one vanishing) while the other can be of type an $I_8$, $I_2^*$, or $IV^*$.
The monodromy of the non-contractible cycles of the curve shows that the $v=0$ monodromy is in fact of $I_8$ type.
Even though it is not captured by a zero of the discriminant, \eqref{Dis45}, the $u=0$ locus is also part of the singular locus since the leading coefficient of the curve is proportional to $u$. Again a careful analysis of the curve identifies this ${\mathfrak S}_u$ singularity as one of $I_2$ type. 

\subsection{$\{4,6\}$}

The allowed $\D_{c_a}$ are
\begin{equation}
\{4,6\}:\quad
\begin{array}{c|c|c|c|c|c|c|c}
 & c_0 & c_1 & c_2 & c_3 & c_4 & c_5 & c_6 \\\hline
\D& 14 & 12 & 10 & 8 & 6 & 4 & 2 \\
\end{array}
\end{equation}
giving the polynomial ansatz 
\begin{equation}\label{ans46}
y^2=u^2 v \l_1+u^3 x \l_2+v^2 x \l_3+u v x^2 \l_4+u^2 x^3 \l_5+v x^4 \l_6+u x^5 \l_7
\end{equation}
in which we can set $\l_7=1$.
The only physical solution of \eqref{intEq} is
\beq
b_3=b_2=b_1=b_0=0,\quad {\rm and} \quad 
-\frac{\l_4}2=\l_5=\l_3,\ 
\l_1=\l_2=0,\ \l_6=-1
\eeq
which, after setting $\l_2=1$, gives 
\beq\label{crv46}
y^2=x\, (u\,x-v)\, (x^3+u\, x-v) ,
\eeq
with discriminant
\beq\label{Dis46}
D_x = -v^{10}(4u^3+27v^2) .
\eeq
This indicates an ${\mathfrak S}_\text{knot}$ singularity of $I_1$ type, and an ${\mathfrak S}_v$ singularity of type $I_4^*$ (after a closer analysis of the homology monodromy of the curve).
As in the previous case, since the leading $x^5$ term in \eqref{crv46} vanishes at $u=0$, the curve degenerates there, and thus there is also an ${\mathfrak S}_u$ singularity.  
This latter singularity can be shown to be of Kodaira type $I_2$.

\begin{table}[t!]
\begin{adjustbox}{center,max width=.9\textwidth}
$\def\arraystretch{1.0}
\begin{array}{c:cc|c:cc|c:c:c:c}
\multicolumn{10}{c}{\Large \textsc{Rank 2 theories with $\varkappa=$1 or 2 and known SW curve}}\\
\hline
\hline
\multicolumn{3}{c|}{ \text{Moduli Space}} &
\multicolumn{3}{c|}{ \text{Flavor and central charges}} 
\\[1mm]
 \D_{u,v}&\ \ d_{\text{HB}}\ \ &\ \  h\ \  
&\quad \ff\quad &\ \ 24a\ \ & 12c 
&\multicolumn{4}{c}{\multirow{-2}{*}{\text{Curve information}}}
\\[1.5mm]
\cdashline{1-10}

\multicolumn{10}{c}{\ef_8-\sof(20)\ \text{series} - I_1\ \text{series}}\\
\hline

 \{6,8\}
& 46 & 0
&\sof(20)_{16} & 202 & 124  
&\multicolumn{4}{c}{\ccg y^2=x\, (u\, x-v)\,(x^4+u\, x-v)}\\
 \{4,10\}
 & 46 & 0
&[\ef_8]_{20} & 202 & 124 
&\multicolumn{4}{c}{y^2=x^5+ (u\, x-v)^3}\\
\{4,6\}
&30 & 0
&\suf(2)_8 \times \sof(16)_{12} & 138 & 84 
&\multicolumn{4}{c}{\ccg y^2=x\, (u\,x-v)\, (x^3+u\, x-v)}\\
 \{4,5\}
&  26 & 0
&\suf(10)_{10} & 122  & 74 
&\multicolumn{4}{c}{\ccg y^2=(u\, x-v)(x^5+u\, x-v)} \\

 \{3,5\}
&  22 & 0
&\sof(14)_{10}\times \uf(1) & 106  & 64 
&\multicolumn{4}{c}{y^2=x(x^5+(u\, x-v)^2) } \\
 \{3,4\}
&18 & 0
&\suf(2)_6\times \suf(8)_8 & 90  & 54  
&\multicolumn{4}{c}{\ccg y^2=(u\, x-v)(x^4+u\, x-v)}\\
\rcy \{2,4\}
& 14 & 0 
& \sof(12)_8 & 74 & 44 
&\multicolumn{4}{c}{y^2=x(x^4+\t\,x^2\,(u\, x-v)+(u\,x-v)^2)}\\
\rcy  \{2,4\} 
& 11 & 1 
& \sof(8)_8\times\suf(2)_5 & 75 & 42  
&\multicolumn{4}{c}{{\rm Degenerate\ genus\ 2\ RS}}\\
\rcy \{2,3\} 
& 10 & 0
& \uf(6)_6 & 58 & 34
&\multicolumn{4}{c}{y^2=x^6+\t x^3\, (u\,x-v)+ (u\, x-v)^2}\\
\rcy \{2,2\}
&  6 & 0
& \suf(2)_4^5 & 42 & 24
&\multicolumn{4}{c}{\ccgy y^2=(u\, x-v)(x^5 + \t_1 x^3 + \t_2 x^2 + \t_3 x + \t_4)}\\

\{\frac32,\frac52\} 
& 6 & 0  
& \suf(5)_5 & 42 & 24 &
\multicolumn{4}{c}{y^2=x^5+(u\,x-v)^2}\\

 \{\frac43,\frac53\}
& 2 & 0
&\suf(2)_{\frac{10}3} \times \uf(1) & 26 & 14 &
\multicolumn{4}{c}{y^2=x(x^5+u\,x-v)}\\

 \{\frac65,\frac85\}
& 1 & 0 &
\suf(2)_{\frac{16}5} & \frac{114}5 & 12 &
\multicolumn{4}{c}{y^2=x(x^4+u\,x-v)}\\
 \{\frac54,\frac32\} 
& 1 & 0
& \uf(1) & 22 & \frac{23}2 &
\multicolumn{4}{c}{y^2=x^6+u\,x-v}\\
 \{\frac87,\frac{10}7\} 
& 0 &0  & \varnothing
& \frac{134}7 & \frac{68}7 &
\multicolumn{4}{c}{y^2=x^5+u\,x-v}\\[.5mm]
\cdashline{1-10}


\multicolumn{10}{c}{\suf(6)\ \text{series} - I_4\ \text{series}}\\
\hline
 \{6,8\} 
 & 23 &1  &
\suf(6)_{16}{\times}\suf(2)_9 & 179 & 101 
&\multicolumn{4}{c}{\ccg y^2=x\, (u\, x-v)\,(x^4+u\, x-v)}\\
 \{4,6\}
& 13 & 1 &
\suf(4)_{12}{\times} \suf(2)_7{\times}\uf(1) & 121 & 67 
&\multicolumn{4}{c}{\ccg y^2=x\, (u\,x-v)\, (x^3 + u\, x-v)}\\
 \{4,5\}
& 11 & 0 &
\suf(3)_{10}{\times} \suf(3)_{10}{\times}\uf(1) & 107 & 59  
&\multicolumn{4}{c}{\ccg y^2=(u\, x-v)(x^5+u\, x-v)}\\
 \{3,5\}
& 8 & 1 &
\suf(3)_{10}{\times} \suf(2)_6{\times}\uf(1) & 92 & 50  
&\multicolumn{4}{c}{y^2=x(x^5+(u\, x-v)^2) }\\
 \{3,4\}
& 6 & 0 &
\suf(2)_8{\times} \suf(2)_8{\times}\uf(1)^2 & 78 & 42 
&\multicolumn{4}{c}{\ccg y^2=(u\, x-v)(x^4+u\, x-v)}\\
\rcy \{2,3\} 
& 2 & 0
& \uf(1)\times \uf(1) & 49 & 25  
&\multicolumn{4}{c}{y^2=x^6+\t x^3\, (u\,x-v)+ (u\, x-v)^2}\\

\cdashline{1-10}


\multicolumn{10}{c}{\spf(14)\ \text{series} - \tilde{I}_4\ \text{series}}\\
\hline
 \{6,8\}
&  29 & 7
&\spf(14)_9 & 185 & 107 
&\multicolumn{4}{c}{\ccg y^2=x\, (u\, x-v)\,(x^4+u\, x-v)}
\\
 \{4,6\}
& 17 & 5
&\suf(2)_8 \times\spf(10)_7& 125  & 71 
&\multicolumn{4}{c}{\ccg y^2=x\, (u\,x-v)\, (x^3 +u\, x-v)}
\\

 \{3,5\}
& 11 & 4
&\spf(8)_6\times \uf(1) & 95  & 53 
&\multicolumn{4}{c}{y^2=x(x^5+(u\, x-v)^2) }
\\
\rcy \{2,4\}
& 6 & 3
& \spf(6)_5 & 65 & 35 
&\multicolumn{4}{c}{y^2=x(x^4+\t\,x^2\,(u\, x-v)+(u\,x-v)^2)}
\\

\cdashline{1-10}

\multicolumn{10}{c}{\suf(5)\ \text{series}\ - I_1^*\ \text{series}}\\
\hline
 \{6,8\}
& 14 & 0
&\suf(5)_{16} & 170 & 92  
&\multicolumn{4}{c}{\ccg y^2=x\, (u\, x-v)\,(x^4+u\, x-v)}\\
 \{4,6\}
& 6 & 0 &
\suf(3)_{12}{\times} \uf(1) & 114 & 60  
&\multicolumn{4}{c}{\ccg y^2=x\, (u\,x-v)\, (x^3 +u\, x-v)}\\
 \{3,5\}
& 3 & 0 &
\suf(2)_{10}{\times} \uf(1) & 86 & 44  
&\multicolumn{4}{c}{y^2=x(x^5+(u\, x-v)^2) }\\[.5mm]
\hline
\hline

\end{array}$
\caption{\label{tab:knownSWr2}
The first 6 columns give the SCFT data while the last column gives the SW curve.
We shade in \green{green} the curves which are derived here and were not known before. 
We shade in \yellow{yellow} $\cN=2$ Lagrangian theories.
}
\end{adjustbox}
\end{table}

\subsection{$\{6,8\}$}\label{sec:68}

The $\D_{c_a}$ are
\begin{equation}
\{6,8\}:\quad
\begin{array}{c|c|c|c|c|c|c|c}
 & c_0 & c_1 & c_2 & c_3 & c_4 & c_5 & c_6 \\\hline
\D& 18 & 16 & 14 & 12 & 10 & 8 & 6 \\
\end{array}
\end{equation}
which lead to the ansatz
\begin{equation}
y^2=u^3 \l_1+v^2 x \l_2+u v x^2 \l_3+u^2 x^3 \l_4+v x^5 \l_5+u x^6 \l_6 .
\end{equation}
Notice that in this case both an $x^6$ and a $x^5$ term is allowed, we have to assume that at least one of the two is not vanishing but cannot yet fix any at a specific value. 
To do a systematic search we will then first assume that $\l_6\neq0$, set it to one and look for physical solutions and then we will repeat the analysis setting $\l_6=0$. 
We find a single solution which passes the discriminant constraints,
\beq
b_3=b_2=b_1=b_0=0,\quad {\rm and} \quad \l_2=\l_4=-\frac{\l_3}2,\ \l_6=-\l_5=-1
\eeq
which, after setting $\l_3=-2$, gives the curve
\beq\label{crv68}
y^2=x\, (u\,x-v)\, (x^4+u\, x-v)
\eeq
with discriminant
\beq\label{Dis68}
D_x = -v^{12}(27u^4+256v^3).
\eeq
This indicates two irreducible components, an unknotted and a knotted one with order of vanishing respectively equal to twelve and one.
The knotted component is of type $I_1$ and the monodromy of a basis of homology cycles of the curve reveals that the unknotted one is of type $I_6^*$.

\section{Current status of rank-2 theories}\label{sec:6}

In this paper we analysed the SK integrability condition for the SW data in canonical frame, and  obtained the complete answer for the case of curves satisfying the polynomial ansatz.
Unfortunately we don't have yet any concrete strategy to extend our systematic analysis to the general case. 
The reason is simple: if we drop the polynomial requirement, scale invariance allows for an infinite set of coefficients for the candidate curve. 
Rather than discussing ideas of how to go about finding the general solution --- some possibilities will be presented in \cite{Argyres:2022fwy} --- we will instead discuss our expectations on how many solutions are left to be found.

\begin{table}[ht]
\begin{adjustbox}{center,max width=1.5\textwidth}
$\def\arraystretch{1.0}
\begin{array}{c:cc|c:cc|c:c:c}
\multicolumn{9}{c}{\large \textsc{Rank 2 theories with $\varkappa=1$ or 2 and no SW curve}}\\
\hline
\hline
\multicolumn{3}{l|}{\quad \text{Moduli Space}} &
\multicolumn{3}{l|}{\quad \text{Flavor and central charges}} &
\multicolumn{3}{c}{\quad \text{Curve information}}\\[1mm]
\D_{u,v}&\ \ d_{\text{HB}}\ \ &\ \  h\ \  
&\quad \ff\quad &\ \ 24a\ \ & 12c 
&\ \ \rSf_{\rm knot}\ \, \, &\ \ \ \rSf_{u}\ \ \, \, &\ \ \ \rSf_{ v}\ \ \, \, 
\\[1.5mm]

\cdashline{1-9}
\multicolumn{9}{c}{\spf(12)-\spf(8)-\ff_4\ \text{series}}\\
\hline
 \{4,6\} 
&  22 & 0
& \spf(12)_8 & 130 & 76  
&(I_1)^2&I_{12}& \varnothing \\
 \{4,6\}
& 20 & 2
&\spf(4)_7 \times \spf(8)_8 & 128 & 74 
&I_1 & I_8 & I_1^* \\
 \{3,4\}
&12 & 0
&\suf(2)_8\times \spf(8)_6 & 84  & 48 
&I_1 & I_8 & I_2 \\
 \{3,4\}
& 11 & 1
&\suf(2)_5\times \spf(6)_6\times \uf(1) & 83  & 47 
&I_1 &I_6 &I_0^*\\
 \{4,5\}
& 16 & 0
&[\ff_4]_{10}\times \uf(1) & 112  & 64
&I_1 &\varnothing & IV^*\\
 \{\frac52,3\}
&7 & 0
& \spf(6)_5\times \uf(1) & 61 & 34
&I_1  & I_6 & \varnothing \\
\rcy \{2,2\}
& 3 & 0
& \spf(4)_4 & 38 & 20 
&(I_1)^4 & I_4 &\varnothing\\
\rcb \{2,2\}
&2 & 2
& \suf(2)_6 & 36 & 18    
&I_0^* & I_0^*& \varnothing \\
\cdashline{1-9}

\multicolumn{9}{c}{\spf(14)\ \text{series}}\\
\hline
\{4,6\}
& 15 & 5
&\suf(2)_5 \times \spf(8)_7 & 123 & 69 
&I_1 & I_0^* &I_2^*
\\
\cdashline{1-9}

\multicolumn{9}{c}{\spf(12)\ \text{series}}\\
\hline
 \{4,10\}
& 32 & 6
&\spf(12)_{11} & 188 & 110
&I_1 & \varnothing &I_5^*
\\
\rcy \{2,4\}
& 8 & 2
&\spf(4)_5 \times\sof(4)_8 & 68 & 38
&(I_2)^2 & \varnothing &I_1^*
\\
\rcy \{2,6\}
& 14 & 4
& \spf(8)_7 & 98 & 56
&(I_1)^2 & \varnothing &I_2^*
\\
 \{\frac43,\frac{10}3\}
& 4 & 2
& \spf(4)_{\frac{13}3} & 48 & 26
&I_1 & \varnothing &I_1^* 
\\
 \{\frac32,\frac52\}
& 2 & 1
& \spf(2)_{7/2}\times\uf(1) & 38 & 20
&I_1 & \varnothing &I_0^* 
\\
[.5mm]
\cdashline{1-9}

\multicolumn{9}{c}{\spf(8)-\suf(2)^2\ \text{series}}\\
\hline
 \{\frac52,4\}
& 2 & 0
&\suf(2)_5 & 67 & 34 
&I_1 & IV &\varnothing\\
\rcb \{2,4\}
&  2 & 
& \suf(2)_{10} & 60 & 30    
&I_0^*&\varnothing& I_0^*\\
\cdashline{1-9}

\multicolumn{9}{c}{\gf_2\ \text{series}}\\
\hline
\{4,6\} 
& 12 & 2
& [\gf_2]_8\times \suf(2)_{14} & 120 & 66    
&I_0^* & I_0^* &\varnothing \\
 \{\frac{10}3,4\}
& 6& 0
&[\gf_2]_{\frac{20}3} & 82  & 44
&I_1 & I_0^* & \varnothing\\
\rcb
\{2,3\}
& 2 & 2
& \suf(2)_8 & 48 & 24  
&I_0^* &\varnothing & \varnothing\\
\cdashline{1-9}

\multicolumn{9}{c}{\textsc{Isolated\ theories}}\\
\hline
 \{4,6\}
&  10 & 4
& \spf(4)_{14}\times\suf(2)_8 & 118 & 64   
&I_1^*  & I_2 & \varnothing\\
\rcb \{2,6\}
& 2 & 2
& \suf(2)_{14} & 84 & 42    
& I_0^* & \varnothing & I_0^*\\
\rcy \{2,4\}
&  0 & 0
& \varnothing & 58 & 28  
& (I_1)^5&\varnothing & I_1\\

\hline\hline
\end{array}$
\caption{\label{tab:r21} The 22 rank 2 theories with $\varkappa=1$ or 2, organised in RG series, for which no SW curve is known. We shade in \yellow{yellow} $\cN=2$ Lagrangian theories and in \blue{blue} $\cN=4$ theories. In the last three columns we list the expected Kodaira type of the singular locus.}
\end{adjustbox}
\end{table}

A way of proceeding is to compare the list of solutions found here --- reported in table \ref{tab:knownSW} --- with the list of known rank-2 theories \cite{Martone:2021drm}. 
To make this comparison we will use results which will be derived systematically in \cite{Argyres:2022puv}. 
In fact, thanks to the techniques discussed there, we are able to compute the SCFT data which are compatible with each given scale invariant CB geometry which can then be matched with the entries in \cite{Martone:2021drm}. In doing that we do not include the new SCFTs found  by our analysis which will instead be discussed in \cite{Argyres:2022puv}. Table \ref{tab:knownSWr2} summarises the results; notice that we have only included the rank-2 theories which have characteristic dimension $\varkappa=\{1,2\}$. 
Table \ref{tab:r21} instead lists all the rank 2 theories in \cite{Martone:2021drm}, again restricting to those which have $\varkappa=\{1,2\}$, which are not accounted for by the polynomial ansatz analysed here.
Since in \cite{Argyres:2022fwy} we show that any SW curve whose generic fibre is not singular can be brought to the ``canonical frame'' considered here, it follows that all the rank 2 geometries in table \ref{tab:r21} are either non-polynomial regular genus 2 curves or are of the ``split" type described in the introduction.
(The restriction on the characteristic dimension only ensures that the geometry is not diagonal but its generic fibre could still be a product of two elliptic curves with different, constant, complex structure.) 

Another, perhaps relevant, piece of information following from our analysis is the observation that for all polynomial solutions, the order of vanishing of the irreducible components of the quantum discriminant \cite{Martone:2020nsy} matches precisely with the order of vanishing of the corresponding Kodaira type of the stratum. 
But this matching fails in the non-polynomial solution presented in \cite{Argyres:2022fwy} corresponding to the curve for the $\cN=4$ $\su(3)$ gauge theory. 
In that case there is mismatch by one: the discriminant has a single knotted component of Kodaira type $I_0^*$, which corresponds in rank 1 to an order of vanishing of the quantum discriminant equal to six, whereas it appears in the curve to vanish at order five instead. 

In table \ref{tab:r21} we have listed the Kodaira type of the various connected components of the singular locus.
Then matching between the expected order of vanishing of the discriminant and the discriminant of the curve can be assessed without finding the actual solution. 
More precisely, scale invariance constrains the scaling dimension of the coefficients of $x^n$ in the curve \eqref{SWcrv} which in turns allows to compute the scaling dimension of the $x$ discriminant regardless of the form of the actual curve.
Doing this exercise is instructive and we find that nearly in all cases listed in table \ref{tab:r21}, the expected discriminant does not match the one from the possible curve, providing some slim evidence that the CB of these SCFTs will be described by a non-polynomial solution. 
Moreover, the way in which the mismatch arises intriguingly seems to have some regularity among curves corresponding to SCFTs which were found in \cite{Martone:2021ixp} to be connected by mass deformations. 

Finally, note that despite the fact that the polynomial ansatz is a strikingly strong restriction on the possible form of the curve, the solutions we find here account for more than half of the known theories with $\varkappa=\{1,2\}$. 
It is realistic to expect that lifting the polynomial constraint, will allow for numerous new solutions and will thus considerably enhance our knowledge of rank 2 SCFTs. 

\acknowledgments 
PCA is supported by DOE grant DE-SC0011784, and MM is supported by STFC grant ST/T000759/1.

\appendix

\section{Remaining solutions}\label{appA}

For completeness we re-derive here the solutions which have already appeared in \cite{Argyres:2005pp, Argyres:2005wx}.

\subsection{$\{\frac87,\frac{10}7\}$}

Using \eqref{SCal} it is straightforward to compute the scaling dimension of the $c_n$ coefficients,
\begin{equation}
\left\{\frac{8}{7},\frac{10}{7}\right\}:\quad
\begin{array}{c|c|c|c|c|c|c|c}
& c_0 & c_1 & c_2 & c_3 & c_4 & c_5 & c_6 \\\hline
\D& \frac{10}{7} & \frac{8}{7} & \frac{6}{7} & \frac{4}{7} & \frac{2}{7} & 0 & -\frac{2}{7} \\
\end{array}
\end{equation}
then the polynomial ansatz acquires the particularly simple form
\begin{equation}
y^2=x^5 \l_3+ +u x \l_2+ v \l_1 .
\end{equation}
Since if $\l_3=0$ the curve is no longer genus 2, we can safely assume it to be non-zero and set it to one. We then find the solution
\beq
b_3=b_2=b_1=b_0=0,\quad {\rm and} \quad \l_1=-\l_2
\eeq
Fixing $\l_2=1$ we are left with the curve,
\beq\label{crv87107}
y^2=x^5+u x -v .
\eeq
The discriminant of the right hand side is
\beq\label{Dis87107}
D_x = 256 u^5+ 3125 v^4 ,
\eeq
from which we determine that $D_x$ only has one irreducible component and the order of vanishing there is equal to one.
Doing the matching between order of vanishing and possible rank 1 geometries, we find that the knotted component is an $I_1$ singularity.

\subsection{$\{\frac65,\frac85\}$}

The scaling dimensions of the $c_a$ in this case are
\begin{equation}
\left\{\frac{6}{5},\frac{8}{5}\right\}:\quad
\begin{array}{c|c|c|c|c|c|c|c}
 & c_0 & c_1 & c_2 & c_3 & c_4 & c_5 & c_6 \\\hline
\D& 2 & \frac{8}{5} & \frac{6}{5} & \frac{4}{5} & \frac{2}{5} & 0 & -\frac{2}{5} \\
\end{array}
\end{equation}
which lead to the polynomial ansatz
\begin{equation}\label{ans6585}
y^2=x(x^4\, \l_3 + u\, x\, \l_2 + v\, \l_1) .
\end{equation}
We can again set $\l_3=1$ without loss of generality. Then plugging \eqref{ans6585} into \eqref{intEq} we find two different solutions but only one for which the linear term in $x$ is not zero for generic values of $(u,v)$. 
This latter is the only physical solution,
\beq
b_3=b_2=b_1=b_0=0,\quad {\rm and} \quad \l_1=-\l_2 .
\eeq
Fixing $\l_2=1$ we are left with the curve
\beq\label{curve6585}
y^2=x(x^4+u x -v), 
\eeq
which has discriminant
\beq\label{Dis6585}
D_x = v^2\,(27 u^4+ 256 v^3)
\eeq
from which we determine that $D_x$ only has one unknotted and one knotted irreducible component with order of vanishing two and one respectively. 
Analysing the monodromies of the curve, we find that the knotted component is an $I_1$ singularity while the $v=0$ component is an $I_2$.

\subsection{$\{\frac54,\frac32\}$}

The scaling dimensions for the $c_a$ are
\begin{equation}
\left\{\frac{5}{4},\frac{3}{2}\right\}:\quad
\begin{array}{c|c|c|c|c|c|c|c}
 & c_0 & c_1 & c_2 & c_3 & c_4 & c_5 & c_6 \\\hline
 \D&\frac{3}{2} & \frac{5}{4} & 1 & \frac{3}{4} & \frac{1}{2} & \frac{1}{4} & 0 \\
\end{array}
\end{equation}
which lead to the polynomial ansatz
\begin{equation}
y^2=v \l_1+u x \l_2+x^6 \l_3 .
\end{equation}
As in the previous cases, we can set $\l_3=1$ after which we find two solutions but only one which is not singular for generic values of $(u,v)$,
\beq
b_3=b_2=b_1=b_0=0,\quad {\rm and} \quad \l_1=-\l_2 .
\eeq
Fixing $\l_2=1$ we are left with the following curve:
\beq\label{crv5432}
y^2=x^6+u x -v ,
\eeq
with discriminant
\beq\label{Dis5432}
D_x = 3125 u^6+ 46656 v^5,
\eeq
implying a single irreducible component with order of vanishing equal to one. 
Doing the matching between order of vanishing and possible rank 1 geometries, we find that the knotted component is an $I_1$ singularity.

\subsection{$\{\frac{4}{3},\frac53\}$}

Using \eqref{SCal} we find that the $\D_{c_a}$ are
\begin{equation}
\left\{\frac{4}{3},\frac{5}{3}\right\}:\quad
\begin{array}{c|c|c|c|c|c|c|c}
 & c_0 & c_1 & c_2 & c_3 & c_4 & c_5 & c_6 \\\hline
\D& 2 & \frac{5}{3} & \frac{4}{3} & 1 & \frac{2}{3} & \frac{1}{3} & 0 \\
\end{array}
\end{equation}
from which we derive the polynomial ansatz
\begin{equation}
y^2=x(v \l_1+u x \l_2+x^5 \l_3) .
\end{equation}
We can set $\l_3=1$ after which we find two solutions but only one which is not singular for generic values of $(u,v)$,
\beq
b_3=b_2=b_1=b_0=0,\quad {\rm and} \quad \l_1=-\l_2 .
\eeq
Fixing $\l_2=1$ we are left with the curve,
\beq\label{crv4353}
y^2=x(x^5+u x -v),
\eeq
with discriminant
\beq\label{Dis4353}
D_x = v^2\,(256 u^5 - 3125 v^4)
\eeq
which indicates that the singular locus has an unknotted and a knotted irreducible component with order of vanishing two and one respectively. 
Analysing the monodromies of the curve, we find that the knotted component is an $I_1$ singularity while the $v=0$ is an $I_2$.

\subsection{$\{\frac32,\frac52\}$}

The last pair of fractional scaling dimensions admitting a solution with polynomial ansatz has $\D_{c_a}$
\begin{equation}
\left\{\frac{3}{2},\frac{5}{2}\right\}:\quad
\begin{array}{c|c|c|c|c|c|c|c}
 & c_0 & c_1 & c_2 & c_3 & c_4 & c_5 & c_6 \\\hline
\D& 5 & 4 & 3 & 2 & 1 & 0 & -1 \\
\end{array} .
\end{equation}
The polynomial ansatz compatible with those values is
\begin{equation}
y^2=v^2 \l_1+u v x \l_2+u^2 x^2 \l_3+x^5 \l_4,
\end{equation}
in which we can set $\l_4=1$ without loss of generality. 
After this we find a unique solution which is also not singular for generic $(u,v)$,
\beq
b_3=b_2=b_1=b_0=0,\quad {\rm and} \quad \l_1=\l_3=-\frac{\l_2}2 .
\eeq
Fixing $\l_2=-2$ we are left with the curve
\beq\label{crv3252}
y^2=x^5+(u\, x- v)^2 ,
\eeq
with discriminant
\beq\label{Dis3252}
D_x = v^5(3125 v^3- 108 u^5) .
\eeq
This expression implies that there are two irreducible components, an unknotted one and a knotted one, with order of vanishing equal to five and one respectively. Doing the matching between order of vanishing and possible rank 1 geometries, we find that the knotted component is an $I_1$ singularity while the unknotted one is an $I_5$.

\subsection{$\{2,2\}$}
\label{appA22}

This case has scaling dimensions equal to two. These are associated with exactly marginal operators of the SCFT, and thus we expect to find a continuous family of solution in this case. 
Since $\D_u=\D_v$, there is a 4-parameter $\GL(2,\C)$ group of linear CB reparametrisations.
In this case $\D_x$ is dimensionless \eqref{scalexy}, and so the $\D_{c_a}$ are all the same,
\begin{equation}
\{2,2\}:\quad
\begin{array}{c|c|c|c|c|c|c|c}
 & c_0 & c_1 & c_2 & c_3 & c_4 & c_5 & c_6 \\\hline
\D& 2 & 2 & 2 & 2 & 2 & 2 & 2 \\
\end{array}
\end{equation}
which implies that the polynomial ansatz involves numerous terms,
\begin{align}
y^2&= (u \l_1+v \l_2) 
+ (u \l_3+v \l_4) x
+ (u \l_5+v \l_6) x^2 
+ (u \l_7+v \l_8) x^3\\
&\quad+ (u \l_9+v \l_{10}) x^4 
+ (u \l_{11}+v \l_{12}) x^5 
+ (u \l_{13}+v \l_{14}) x^6 .
\end{align}
Since there are both $x^6$ and $x^5$ terms, we can break the analysis into cases.
Assuming that a coefficient of the $x^6$ terms does not vanish, then we can use two parameter degrees of freedom of the $\GL(2,\C)$ reparameterizations to set the coefficient of $x^6$ to $u$, \emph{i.e.}, we set $\l_{14}=0$ and $\l_{13}=1$.
Then we can use the remaining two parameters to the coefficient of one of the next powers of $x$ to $v$.
This gives a single family of solutions
\begin{align}\label{22lambsol}
b_3&=b_2=b_1=b_0=0,\quad  
\l_{14}= \l_{11}=\l_{10}=\l_9=\l_8=\l_1=0, \nn\\ 
& \l_{13}= -\l_{12}=1, \ {\rm and} \quad 
\l_6= -\l_7, \
\l_4= -\l_5, \ 
\l_2= -\l_3 ,
\end{align}
with four coefficients unfixed.
If, instead, the coefficients of $x^6$ vanish, then we can set the coefficient of $x^5$ to $u$, and, repeating the same procedure we find a 3-parameter family of solutions after fixing the CB reparametrisation freedom.  However, this curve is just a specialization of the 4-parameter \eqref{22lambsol}.
The curve corresponding to \eqref{22lambsol} is
\beq
y^2=(u\, x-v)(x^5 + \t_1 x^3 + \t_2 x^2 + \t_3 x + \t_4)
\eeq
with discriminant
\beq\label{Dis22}
D_x = \prod_{i=1}^5(u+\a_i v)^2
\eeq
with $\a_i$ which are functions of the $\t_i$. This implies that we have five irreducible components with order of vanishing equal to two. Since the scaling dimension of $u$ and $v$ are equal in this case, there is no intrinsic notion of knotted vs.\ unknotted. By analysing the curve in more depth we lift the possible ambiguity between a type $I_2$ and a type $II$ singularity, establishing that all five irreducible components are $I_2$'s. 

This solution has an unexpected number of free parameters.  Since there are only two CB operators of dimension 2, one expects from the usual representation theory arguments that a unitary $\cN{=}2$ SCFT should have just two exactly marginal deformations, and so only a 2-parameter family of CB geometries.
We do not know what, if any, feature of these CB geometries picks out a special 2-parameter subfamily as the one corresponding to the physical theory.

Note that these parameters cannot be removed by our reparametrisation freedoms.
It is easiest to see this by going to projective $(x,w)$ coordinates where the curve reads
\beq
y^2=(u\, x-v\, w) (x^5 + \t_1 x^3 w^2 + \t_2 x^2 w^3 
+ \t_3 x w^4 + \t_4 w^5) .
\eeq
In particular, a $\GL(2,\C)$ reparametrisation of $(x,w)$ together with the associated one of $(u,v)$  which leaves the $(ux-vw)$ factor invariant, leaves the factorized form of the curve invariant, but can be used to change the coefficients of the degree-5 factor, $x^5 + \t_1 x^3 w^2 + \t_2 x^2 w^3 + \t_3 x w^4 + \t_4 w^5$.
Keeping it monic and the coefficient of $x^4$ zero, leaves two parameters, which can be used to set, say, $\t_3$ and $\t_4$ to fixed values.
But this does not reduce the number of parameters describing the CB geometry, since such a reparametrization changes the holomorphic 1-form basis from the canonical frame to one which depends on $\t_3$ and $\t_4$.

\subsection{$\{2,3\}$}

We now proceed to another pair of scaling dimensions with integer entries. Because of the presence of a CB scaling dimension with dimension two we expect to find a one parameter family of solutions, and since the degrees of the $\SU(3)$ Casimirs are $(2,3)$, we expect to find a set of SCFT data compatible with the two inequivalent $SU(3)$ $\cN=2$ SCFTs. 
Namely $\SU(3)$ with hypermultiplets either in  $6\,{\bf 3}$ or ${\bf 3}\oplus {\bf 6}$.  
Start by tabulating the $\D_{c_a}$,
\begin{equation}
\{2,3\}:\quad
\begin{array}{c|c|c|c|c|c|c|c}
 & c_0 & c_1 & c_2 & c_3 & c_4 & c_5 & c_6 \\\hline
\D& 6 & 5 & 4 & 3 & 2 & 1 & 0 \\
\end{array}
\end{equation}
which lead to the following polynomial ansatz
\begin{equation}
y^2=u^3 \l_1+v^2 \l_2+u v x \l_3+u^2 x^2 \l_4+v x^3 \l_5+u x^4 \l_6+x^6 \l_7 .
\end{equation}
Since there is a single term with order greater than four in $x$, we can then set its coefficient to one without loss of generality: $\l_7=1$. 
Despite fixing this coefficient, we find various solutions of \eqref{intEq}. 
As usual there are many solutions which are singular for generic $(u,v)$ (for which the $x$ discriminant of the curve vanishes identically) and have to then be discarded. 
There is also one solution for which the $x$ discriminant has a pole at $v=0$; this is also unphysical. 
We are then left with the only solution
\beq
b_3=b_2=b_1=b_0=0,\quad {\rm and} \quad \l_2=\l_4=-\frac{\l_3}2,\ \l_6=-\l_5 .
\eeq
Fixing $\l_3=-2$ we are left with a one-parameter families of curves. As in the previous case, the free parameter $\l_5$ is expected because of the presence of a CB scaling dimension two parameter and ought to be identified with the holomorphic gauge coupling of the $\SU(3)$ gauge group.
Thus we rename it as $\l_5=\t$, giving the curve
\beq\label{crv23}
y^2=x^6+x^3\, (u\,x-v)\t + (u\, x-v)^2 ,
\eeq
with discriminant
\beq\label{Dis23}
D_x \propto v^6 (u^3+ \a_1 v^2)(u^3+ \a_2 v^2)
\eeq
where $\a_1$ and $\a_2$ depend non-trivially on $\t$. This expression implies that there are three irreducible components, an unknotted and two knotted ones, with order of vanishing equal to six, one, and one, respectively. Doing the matching between order of vanishing and possible rank-1 geometries, we find that the two knotted components are $I_1$'s while, after studying the behaviour of the non-trivial cycles to lift the ambiguity between an $I_6$ and $I_0^*$, we establish that the unknotted component is an $I_6$.

\subsection{$\{2,4\}$}

This is another pair of scaling dimensions with one entry equal to $2$, we thus expect to find a one-parameter family of solutions. 
The $\D_{c_a}$'s are
\begin{equation}
\{2,4\}:\quad
\begin{array}{c|c|c|c|c|c|c|c}
 & c_0 & c_1 & c_2 & c_3 & c_4 & c_5 & c_6 \\\hline
\D& 10 & 8 & 6 & 4 & 2 & 0 & -2 \\
\end{array} .
\end{equation}
These allow for many terms in the polynomial ansatz,
\begin{align}
y^2&=u^5 \l_1+u^3 v \l_2+u v^2 \l_3+u^4 x \l_4+u^2 v x \l_5+v^2 x \l_6\,+\\\nonumber
&\qquad u^3 x^2 \l_7+ u v x^2 \lambda_8+u^2 x^3 \l_9+v x^3 \l_{10}+u x^4 \l_{11}+x^5 \l_{12} ,
\end{align}
but since there is a unique $x^5$ coefficient and no $x^6$, we can assume it to be non-zero and set it to one. 
Then the integrability equation gives two physical solution but which, using the fact that $\D_u|\D_v$, which enhances our reparametrisation freedom, can actually non-trivially mapped into one another. 
We are thus left with only one inequivalent physical solution,
\begin{align}\nonumber
b_3=b_2=&\,b_1=b_0=0,\quad  \l_1=\l_2=\l_3=\l_4=\l_5=\l_7=0\\
&{\rm and} \quad \l_6=\l_9=-\frac{\l_8}2,\ \l_{10}=-\l_{11} .
\end{align}
Setting $\l_8=-2$ we are left with the following one-parameter families of curves where we rename $\l_{11}=\t$, which ought to be identified with the holomorphic gauge coupling of the $\Sp(4)$ gauge group,
\beq\label{crv24}
y^2=x \big(x^4+\t x^2(u\, x-v)+(u\, x-v)^2 \big) .
\eeq
This has discriminant
\beq\label{Dis24}
D_x \propto v^8(u^2+\a_1\, v)(u^2+\a_2\, v),
\eeq
where $\a_1$ and $\a_2$ depend non-trivially on $\t$. This expression implies that there are three irreducible components, an unknotted and two knotted ones, with order of vanishing equal to eight, one, and one, respectively. Analysing the curve in more depth we find that the two knotted components are $I_1$'s while the unknotted one is an $I_2^*$ lifting the ambiguity between an $I_{8}$, an $I_2^*$, and a $IV^*$ from the sole knowledge of the order of vanishing of the discriminant.

\subsection{$\{3,5\}$}

This pair of scaling dimensions does not have any entry equal to two, thus we again expect to find a single solution once we fix all our reparametrisation freedom. The $\D_{c_a}$ are again all integers,
\begin{equation}
\{3,5\}:\quad
\begin{array}{c|c|c|c|c|c|c|c}
 & c_0 & c_1 & c_2 & c_3 & c_4 & c_5 & c_6 \\\hline
\D& 12 & 10 & 8 & 6 & 4 & 2 & 0 \\
\end{array}
\end{equation}
leading to the following polynomial ansatz
\begin{equation}
y^2=u^4 \l_1+v^2 x \l_2+u v x^2 \l_3+u^2 x^3 \l_4+x^6 \l_5 .
\end{equation}
We can as usual set the leading coefficient $\l_5$ to one after which \eqref{intEq} leads to a single solution with discriminant either not identically zero or with poles at finite $(u,v)$ values,
\beq
b_3=b_2=b_1=b_0=0,\quad {\rm and} \quad -\frac{\l_3}2=\l_4=\l_2,\ \l_1=0 .
\eeq
After setting $\l_2=1$ this gives the curve
\beq\label{crv35}
y^2=x\, \big(x^5+(u\,x-v)^2\big) ,
\eeq
with discriminant
\beq\label{Dis35}
D_x = v^9(108u^5+ 3125 v^3) .
\eeq
Thus the singular locus of \eqref{crv35} has two irreducible components, an unknotted and a knotted one with order of vanishing respectively equal to nine and one. Matching the order of vanishing with possible rank 1 geometries, we find that the knotted component is an $I_1$ while, after studying the behaviour of the non-trivial cycles to lift the ambiguity between an $I_9$, an $I_3^*$ and a $III^*$, we establish that the unknotted one is an $I^*_3$.

\subsection{$\{4,10\}$}

In this case the allowed $\D_{c_a}$ are
\begin{equation}
\{4,10\}:\quad
\begin{array}{c|c|c|c|c|c|c|c}
 & c_0 & c_1 & c_2 & c_3 & c_4 & c_5 & c_6 \\\hline
\D& 30 & 24 & 18 & 12 & 6 & 0 & -6 \\
\end{array}
\end{equation}
which leads to the ansatz
\begin{equation}
y^2=u^5 v \l_1+v^3 \l_2+u^6 x \l_3+u v^2 x \l_4+u^2 v x^2 \l_5+u^3 x^3 \l_6+x^5 \l_7
\end{equation}
in which can safely set the leading coefficient $\l_7$ to one. Then the integrability equation leads to only one physical solution,
\beq
b_3=b_2=b_1=b_0=0,\quad {\rm and} \quad \l_4=-\l_5=-3\l_2=3\l_6,\ \l_1=\l_3=0,\ \l_6=-1 ,
\eeq
which gives the following curve after setting $\l_6=1$,
\beq\label{crv410}
y^2=x^5+(u\,x-v)^3 ,
\eeq
with discriminant
\beq\label{Dis410}
D_x = v^{10}(108u^5+3125v^2) .
\eeq
This indicates two irreducible components, an unknotted and a knotted one with order of vanishing respectively equal to ten and one. Analysing the curve further we find that the knotted component is an $I_1$ while the unknotted one is a $II^*$ which lifts the ambiguity of the interpretation given just analysing the order of vanishing of \eqref{Dis410} between an $I_{10}$, an $I_4^*$ and the actual $II^*$. 

\bibliographystyle{Auxiliary/JHEP}

\end{document}